\documentclass[aps,prb,reprint,groupedaddress]{revtex4-2}

\usepackage{graphicx}
\usepackage{amsmath,amssymb}

\begin{document}




\title{Significant Contributions of the Higgs Mode and Self-Energy Corrections to Low-Frequency Complex Conductivity in DC-Biased Superconducting Devices}


\author{Takayuki Kubo}
\email[]{kubotaka@post.kek.jp}
\affiliation{High Energy Accelerator Research Organization (KEK), Tsukuba, Ibaraki 305-0801, Japan}
\affiliation{The Graduate University for Advanced Studies (Sokendai), Hayama, Kanagawa 240-0193, Japan}



\begin{abstract}
We investigate the complex conductivity of superconductors under a direct current (DC) bias based on the well-established Keldysh-Eilenberger formalism of nonequilibrium superconductivity. This robust framework allows us to account for the Higgs mode and impurity scattering self-energy corrections, which are now known to significantly impact the complex conductivity under a bias DC, especially near the resonance frequency of the Higgs mode. The purpose of this paper is to explore the effects of these contributions on the low-frequency complex conductivity relevant to superconducting device technologies.
We begin by nonperturbatively calculating the equilibrium Green's functions under a bias DC, followed by an analysis of the time-dependent perturbative components. This approach enables us to derive the complex conductivity formula for superconductors ranging from clean to dirty limits, applicable to any bias DC strength. We validate our theoretical approach by reproducing known results and experimentally observed features, such as the characteristic peak in \(\sigma_1\) and the dip in \(\sigma_2\) attributed to the Higgs mode. Our calculations reveal that the Higgs mode and impurity scattering self-energy corrections significantly affect the complex conductivity even at low frequencies (\(\hbar \omega \ll \Delta\)), relevant to superconducting device technologies.
Specifically, we find that the real part of the low-frequency complex conductivity, \(\sigma_1\), exhibits a bias-dependent reduction up to \(\hbar \omega \sim 0.1\), a much higher frequency than previously considered. This finding allows for the suppression of dissipation in devices by tuning the bias DC. Additionally, through the calculation of the imaginary part of the complex conductivity, \(\sigma_2\), we evaluate the bias-dependent kinetic inductance for superconductors ranging from clean to dirty limits. The bias dependence becomes stronger as the mean free path decreases. Our dirty-limit results coincide with previous studies based on the oscillating superfluid density (the so-called slow experiment) scenario. This widely used scenario can be understood as a phenomenological implementation of the Higgs mode into the kinetic inductance calculation, now justified by our calculation based on the robust theory of nonequilibrium superconductivity, which microscopically treats the Higgs mode contribution.
These results highlight the importance of considering the Higgs mode and impurity scattering self-energy corrections in the design and optimization of superconducting devices under a bias DC.
\end{abstract}

\maketitle


\section{Introduction} \label{introduction}

The interaction between external fields and superconducting materials not only unveils fundamental properties of superconductors but also has significant technological implications. Specifically, the superposition of electromagnetic fields on direct current (DC) bias introduces complex dynamics that are of interest from both fundamental~\cite{Moor, Nakamura, Jujo, Shimano_review} and practical perspectives~\cite{2012_Zmuidzinas, 2017_Gurevich_review}.

Recent advancements have significantly enhanced our understanding of the electromagnetic response of superconductors under DC bias, particularly highlighting the critical role of the Higgs mode~\cite{Moor, Nakamura, Jujo, Shimano_review}: oscillations in the amplitude of a superconductor's order parameter~\cite{Shimano_review, Tsuji_review, Anderson}. In the absence of a DC current, interactions with the Higgs mode are confined to terms quadratic in the electromagnetic field~\cite{Shimano_review, Tsuji_review, Tsuji_Aoki, 2013_Matsunaga, 2014_Matsunaga, 2018_Jujo, Silaev}. However, Moor et al.~\cite{Moor} theoretically demonstrated through perturbative calculations that a finite DC supercurrent enables linear coupling of the electromagnetic field to the Higgs mode, a phenomenon that was later experimentally confirmed~\cite{Nakamura}. This research primarily focused on admittance calculations for superconductors in the dirty limit under DC bias, emphasizing how such conditions amplify the visibility of the Higgs mode's effects within the linear response regime.
Building on this foundational research, Jujo~\cite{Jujo} developed a more detailed formulation to calculate surface resistance, applicable across various impurity concentrations and any DC bias intensity. His findings reveal that the impact of the Higgs mode and the impurity scattering self-energy corrections on surface resistance is significantly influenced by both the intensity of the DC bias and the levels of impurities, leading to a diverse range of behaviors under different DC bias conditions.

On the other hand, from a technological perspective, there is significant interest in modulating low-frequency (\(\hbar \omega \ll \Delta\)) complex conductivity using a DC bias. In this context, Gurevich~\cite{2014_Gurevich} demonstrated the non-monotonic behavior of the real part of low-frequency complex conductivity, \(\sigma_1\), in dirty limit superconductors subjected to DC bias. Specifically, as the DC bias increases, \(\sigma_1\) initially decreases, reaching a minimum before increasing again. This behavior indicates that the surface resistance or the quality factor of a resonator could be tuned using a DC bias~\cite{2017_Gurevich_review, 2014_Gurevich, 2020_Kubo_1}. However, the response of \(\sigma_1\) in clean superconductors under similar conditions remains unexplored. Furthermore, the potential effects of the Higgs mode and self-energy corrections on the low-frequency \(\sigma_1\) in both clean and dirty superconductors have yet to be thoroughly investigated.

Additionally, the imaginary part of low-frequency complex conductivity, \(\sigma_2\), plays a pivotal role in superconducting devices~\cite{2012_Zmuidzinas}. For instance, frequency shifts in resonators due to DC bias are influenced by changes in \(\sigma_2\) or kinetic inductance (\(L_k\)), where \(\delta f_r/f_r \propto \delta \sigma_2/\sigma_2 \propto \delta L_k/L_k\) (see e.g., Refs.~\cite{Clem_Kogan, 2020_Semenov, 2020_Kubo_2}). Similar to \(\sigma_1\), the response of \(\sigma_2\) or the kinetic inductance in clean superconductors under comparable conditions remains unexplored. Furthermore, the potential impacts of the Higgs mode and self-energy corrections on the bias-dependent kinetic inductance in both clean and dirty superconductors have yet to be addressed.

As detailed in Ref.~\cite{Rainer_Sauls}, the complex conductivity formula generally includes vertex corrections, namely, the Higgs mode and the impurity scattering self-energy corrections. These corrections vanish in the zero current state under certain assumptions. However, they are known to significantly impact complex conductivity under a bias DC, especially near the resonance frequency of the Higgs mode, and cannot be disregarded~\cite{Moor, Jujo}, as demonstrated by the experimental observation of the Higgs mode in a biased superconductor~\cite{Nakamura}. This situation holds not only around the resonance frequency of the Higgs mode ($\simeq 2\Delta$) but also at lower frequencies (\(\ll \Delta\)), which are highly relevant to superconducting device technologies but remain unexplored.

Our study aims to bridge these gaps by exploring the complex conductivity of DC biased superconductors across the spectrum from clean to dirty limits and from lower (\(\ll \Delta\)) to higher ($\simeq 2\Delta$) frequencies, taking the Higgs mode and impurity scattering self-energy corrections into account and focusing on their implications for the performance of superconducting devices. Our calculations employ the well-established Keldysh-Eilenberger formalism of nonequilibrium superconductivity~\cite{Kopnin, Rainer_Sauls} and assume that the penetration depth \(\lambda\) is much larger than the coherence length \(\xi\), enabling the superconductor to adhere to local electrodynamics. This robust theoretical framework allows us to understand the impact of both the Higgs mode and impurity scattering self-energy corrections on the low-frequency complex conductivity, providing valuable insights for the optimization of superconducting device technologies under a DC bias.

The structure of this paper is outlined as follows:
Section II provides an overview of the Keldysh-Eilenberger formalism for nonequilibrium superconductivity.
In Section III, we calculate the equilibrium Green's functions under a bias DC, including evaluations of the pair potential, the depairing current density, and the quasiparticle spectrum. This section also addresses the calculation of nonequilibrium Green's functions under a time-dependent electromagnetic perturbation superimposed on the bias DC.
In Section IV, we derive the formula for complex conductivity and evaluate it across various frequencies and strengths of bias DC.
Section V discusses the broader implications of our findings for superconducting technologies and outlines future research directions.

\section{Theory}

\begin{figure}[tb]
   \begin{center}
   \includegraphics[width=0.98\linewidth]{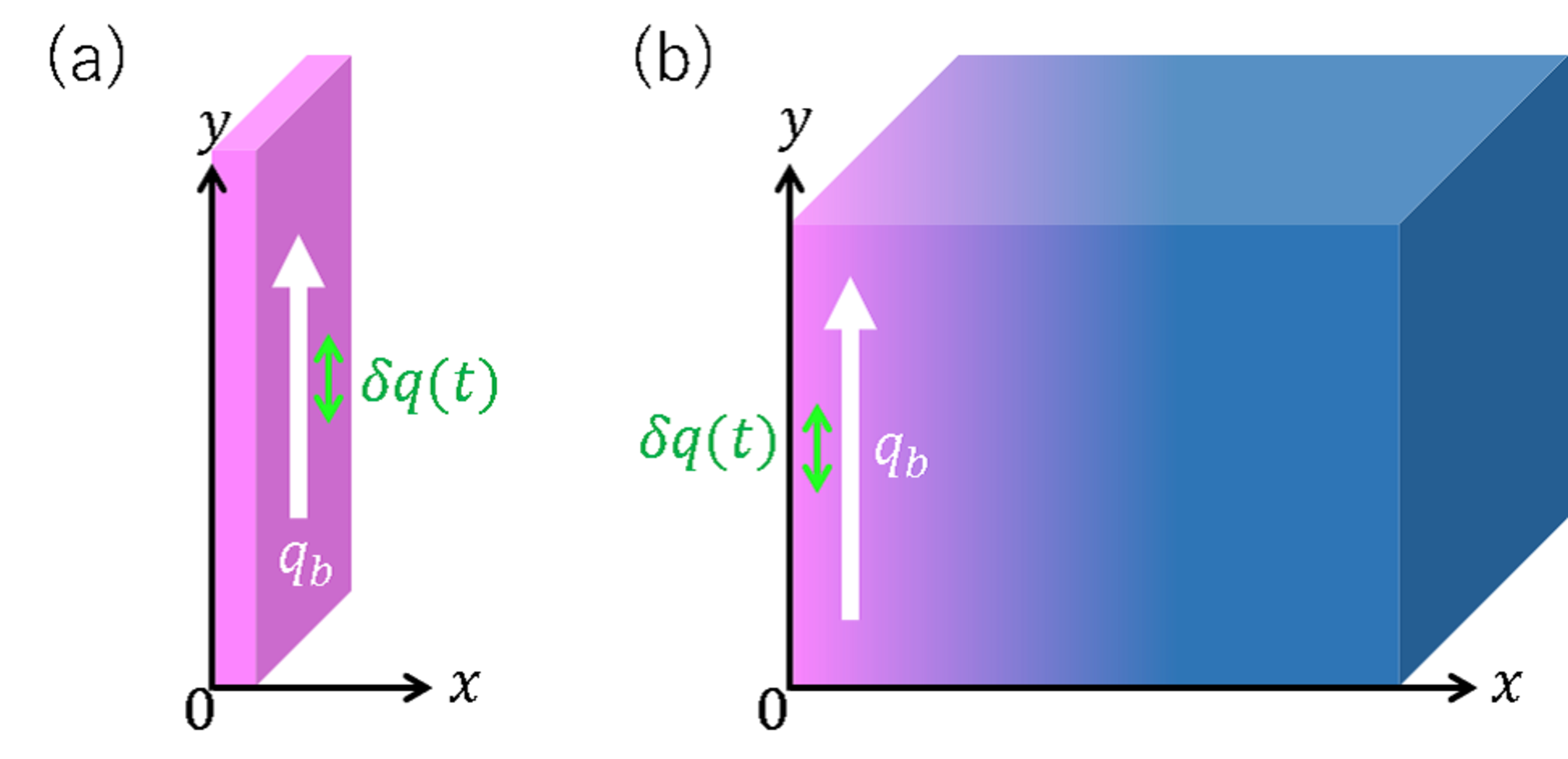}
   \end{center}\vspace{0 cm}
   \caption{
Geometries considered in this study. (a) A thin film and (b) a bulk superconductor, each subjected to a bias DC of arbitrary strength. A weak electromagnetic field is superimposed on the DC bias in both configurations. The constant bias due to the DC component is represented by \(q_b\) (bias superfluid momentum), and the time-dependent perturbation, considered parallel to \(q_b\) for simplicity, is denoted by \(\delta q(t)\). In a bulk superconductor, both \(q_b\) and \(\delta q\) vary with depth \(x\) due to the Meissner effect.
   }\label{fig1}
\end{figure}

\subsection{Nonequilibrium quasiclassical Green's functions}

As depicted in Figure 1, we analyze both a thin-film and a semi-infinite superconductor. 
A bias DC of arbitrary strength flows parallel to the $y$-axis, with a weak electromagnetic field superimposed onto it. 
We employ the Keldysh-Eilenberger quasiclassical Green's functions formalism of nonequilibrium superconductivity (see e.g., Refs.~\cite{Kopnin, Rainer_Sauls}).

Consider a thin film as shown in Fig.~\ref{fig1}(a). In this scenario, we assume a uniform current distribution. The spatial-differentiation terms manifest solely through the phase gradient, $\nabla \chi$, which combines with the electromagnetic vector potential, $\mathbf{A}$, into the superfluid momentum, $\mathbf{q} = \nabla \chi - (2e/\hbar) \mathbf{A}$. Under these conditions, the Eilenberger equations describing the current-carrying state are specified as follows: 
\begin{eqnarray}
\Bigl[ \Bigl( \epsilon - \frac{1}{2}\hbar {\bf v}_f \cdot {\bf q}\Bigr) \hat{\tau}_3 - \hat{\sigma}^r , \hat{g}^r \Bigr]_{\circ} =0 , \label{EilenbergerRA1} 
\end{eqnarray}
and
\begin{eqnarray}
&&\biggl\{  \Bigl( \epsilon -\frac{1}{2} \hbar {\bf v}_f \cdot {\bf q} \Bigr) \hat{\tau}_3 - \hat{\sigma}^R \biggr\} \circ \hat{g}^K + \hat{g}^R \circ \hat{\sigma}^K \nonumber \\
&&- \hat{g}^K \circ \biggl\{ \Bigl( \epsilon -\frac{1}{2} \hbar {\bf v}_f \cdot {\bf q} \Bigr) \hat{\tau}_3 - \hat{\sigma}^A \biggr\}  - \hat{\sigma}^K \circ \hat{g}^A   =0 .\label{EilenbergerK1}
\end{eqnarray}
Here, \(\hat{g}^r (\epsilon, t)\) (where \(r = R, A\)) denotes the quasiclassical retarded and advanced Green's functions, and \(\hat{g}^K(\epsilon, t)\) represents the Keldysh component of the quasiclassical Green's function. 
The commutator \([\alpha, \beta]_{\circ}\) is defined as \(\alpha \circ \beta - \beta \circ \alpha \), and the circle product \(( \alpha \circ \beta ) (\epsilon, t)\) is given by $(\alpha \circ \beta) (\epsilon, t) = \exp[ (i\hbar/2) (\partial_{\epsilon}^\alpha \partial_{t}^\beta - \partial_{t}^\alpha \partial_{\epsilon}^\beta) ] \alpha(\epsilon,  t) \beta(\epsilon, t)$. 
\(\mathbf{v}_f\) represents the Fermi velocity. 
The self-energy $\hat{\sigma}=\hat{\sigma}_{\rm imp}-\hat{\Delta}$ comprises the impurity self-energy \(\hat{\sigma}_{\rm imp}\) and the gap function \(\hat{\Delta}\), 
where $\hat{\Delta} = i\hat{\tau}_2 \Delta$, 
$\hat{g}^R= \hat{\tau}_3 g +  i\hat{\tau}_2 f $, 
and $\hat{g}^A=-\hat{\tau}_3\hat{g}^{R\dagger} \hat{\tau}_3$. 
The Green's functions fulfill the normalization conditions \(\hat{g}^r \circ \hat{g}^r = \hat{1}\) and $\hat{g}^R \circ \hat{g}^K + \hat{g}^K \circ \hat{g}^A =0$.

Consider the Eilenberger equation for the retarded (R) and advanced (A) components, as outlined in Eq.~(\ref{EilenbergerRA1}). We define \(\mathbf{q} = \mathbf{q}_b + \delta \mathbf{q}(t)\), where \(\mathbf{q}_b\) represents the constant bias due to the DC component, and \(\delta \mathbf{q}(t)\) denotes the time-dependent perturbation. For simplicity in this analysis, \(\delta \mathbf{q}(t)\) is assumed to be parallel to \(\mathbf{q}_b\).
Consequently, the Green's function \(\hat{g}^r(\epsilon, t)\) can be expressed as \(\hat{g}^r_b(\epsilon) + \delta \hat{g}^r(\epsilon, t)\), and the self-energy is written as \(\hat{\sigma}^r(\epsilon, t) = \hat{\sigma}^r_b(\epsilon) + \delta \hat{\sigma}^r(\epsilon, t)\). Here, \(\hat{\sigma}^r_b(\epsilon) = -i\gamma \langle \hat{g}_b^r \rangle - \hat{\Delta}_b\), and \(\delta \hat{\sigma}^r(\epsilon, t) = \delta \hat{\sigma}^r_{\rm imp}(\epsilon, t) - \delta \hat{\Delta}(t)\). The angular average of a quantity \(X\) over the Fermi surface is denoted by \(\langle X \rangle\). 
Assuming a spherical Fermi surface, the angular average \(\langle X \rangle\) is calculated using \(\langle X \rangle = \frac{1}{2}\int_0^{\pi} X \sin \theta \, d\theta\), 
where $\theta =\cos^{-1} ({\bf v}_f\cdot {\bf q}_b/ v_f q_b)$. 
The ratio \(\gamma / \Delta_0 = \pi \xi_0/2 \ell\) represents the impurity scattering rate, where \(\xi_0\) is the BCS coherence length, \(\Delta_0\) represents the superconducting gap in the zero-current state at zero temperature, and \(\ell\) is the mean free path of electrons between collisions with nonmagnetic impurities. This decomposition enables the reformulation of Eq.~(\ref{EilenbergerRA1}) as follows:
\begin{eqnarray}
\Bigl[ \bigl( \epsilon - W_b \cos\theta \bigr) \hat{\tau}_3 - \hat{\sigma}^r_b(\epsilon)   , \hat{g}_b^r(\epsilon) \Bigr] =0 , \label{Eilenberger_RA2}   
\end{eqnarray}
and
\begin{eqnarray}
&&\Bigl[ \bigl( \epsilon - W_b \cos\theta \bigr) \hat{\tau}_3 - \hat{\sigma}^r_b(\epsilon) , \delta \hat{g}^r(\epsilon,t) \Bigr]_{\circ} \nonumber \\
&&- \bigl[ \delta W(t) \cos\theta \hat{\tau}_3 + \delta \hat{\sigma}^r(\epsilon,t) , \hat{g}_b^r(\epsilon) \bigr]_{\circ}=0 . \label{Eilenberger_RA3}  
\end{eqnarray}
Here, $W_b = (\hbar/2) v_f q_b$, $\delta W = (\hbar/2) v_f  \delta q$, and ${\bf q}_b / \! / {\bf \delta q}$. 
The equations correspond to the equilibrium state and the time-dependent perturbation, respectively. Specifically, the normalization condition for the retarded (R) and advanced (A) components is expressed as \(\hat{g}_b^r \hat{g}_b^r = \hat{1}\) for the equilibrium state, while for the time-dependent perturbation, it is given by \(\hat{g}_b^r \circ \delta \hat{g}^r + \delta \hat{g}^r \circ \hat{g}_b^r = 0\).

Similarly, we decompose the Keldysh components as follows: \(\hat{g}^{K} = \hat{g}_b^{K} + \delta \hat{g}^{K}\) and \(\hat{\sigma}^{K} = \hat{\sigma}_b^{K} + \delta \hat{\sigma}^{K}\). 
Consequently, the Eilenberger equation for the Keldysh component, as specified in Eq.~(\ref{EilenbergerK1}), can be rewritten to incorporate these decompositions. 
For the equilibrium parts, we have
\begin{eqnarray}
&&\Bigl\{ \bigl( \epsilon -W_b\cos\theta \bigr) \hat{\tau}_3 - \hat{\sigma}_b^R(\epsilon) \Bigr\} \hat{g}_b^K(\epsilon) \nonumber \\
&&- \hat{g}_b^K(\epsilon) \Bigl\{ \bigl( \epsilon -W_b\cos\theta \bigr) \hat{\tau}_3 - \hat{\sigma}_b^A(\epsilon) \Bigr\} \nonumber \\
&& + \hat{g}_b^R(\epsilon) \hat{\sigma}_b^K(\epsilon)  - \hat{\sigma}_b^K(\epsilon) \hat{g}_b^A(\epsilon)   =0 , \label{Eilenberger_K2}
\end{eqnarray}
whose solutions are
\begin{eqnarray}
\hat{g}_b^K &=& (\hat{g}_b^R - \hat{g}_b^A ) \tanh \frac{\epsilon}{2kT} , \label{Eilenberger_K2_g} \\
\hat{\sigma}_b^K &=& (\hat{\sigma}_b^R - \hat{\sigma}_b^A ) \tanh \frac{\epsilon}{2kT} .  \label{Eilenberger_K2_sigma} 
\end{eqnarray}
For the perturbative parts, we have
\begin{eqnarray}
&&\Bigl\{ \bigl( \epsilon -W_b\cos\theta \bigr) \hat{\tau}_3 - \hat{\sigma}_b^R(\epsilon) \Bigr\} \circ \delta \hat{g}^K (\epsilon, t) \nonumber \\
&&- \delta \hat{g}^K(\epsilon, t) \circ \Bigl\{ \bigl( \epsilon -W_b\cos\theta \bigr) \hat{\tau}_3 - \hat{\sigma}_b^A(\epsilon) \Bigr\}   \nonumber \\
&&+ \hat{g}_b^R(\epsilon) \circ \delta \hat{\sigma}^K(\epsilon, t) - \delta \hat{\sigma}^K(\epsilon, t) \circ \hat{g}_b^A(\epsilon) \nonumber \\
&& + \hat{g}_b^{K}(\epsilon) \circ \bigl\{ \delta W(t) \cos\theta\hat{\tau}_3 + \delta \hat{\sigma}^A(\epsilon, t) \bigr\}  \nonumber \\
&& - \bigl\{ \delta W(t) \cos\theta \hat{\tau}_3 + \delta \hat{\sigma}^R(\epsilon, t) \bigr\} \circ \hat{g}_b^{K}(\epsilon) \nonumber \\
&&-\hat{\sigma}_b^K(\epsilon) \circ \delta \hat{g}^A(\epsilon, t) + \delta \hat{g}^R(\epsilon, t) \circ \hat{\sigma}_b^K(\epsilon)
=0 .\label{Eilenberger_K3}
\end{eqnarray}
The normalization conditions become $\hat{g}_b^{R} \hat{g}_b^{K} + \hat{g}_b^{K}\hat{g}_b^{A}=0$ and $\hat{g}_b^{R}\circ \delta \hat{g}^{K} + \delta \hat{g}^{K}\circ \hat{g}_b^{A} + \delta \hat{g}^{R} \circ \hat{g}_b^{K} + \hat{g}_b^{K} \circ \delta \hat{g}^{A}=0$.

The Eilenberger equations are coupled with the gap equation, expressed as
\begin{eqnarray}
\Delta = -\frac{g}{8}\int d\epsilon \Bigl\langle {\rm Tr} [(\tau_1-i\tau_2) \hat{g}^K] \Bigr\rangle , \label{gap_equation1}
\end{eqnarray}
where $g$ is the BCS coupling constant. 
From this, we derive the expressions for the equilibrium and perturbed components of the gap equation:
\begin{eqnarray}
&&\Delta_b = \frac{g}{2} \int d\epsilon  {\rm Re} \langle f_b \rangle \tanh \frac{\epsilon}{2kT} , \label{gap_equation2} \\
&&\delta\Delta = -\frac{g}{8}\int d\epsilon \Bigl\langle {\rm Tr} [(\tau_1-i\tau_2) \delta \hat{g}^K] \Bigr\rangle . \label{gap_equation3}
\end{eqnarray}

In summary, our system is governed by Eqs.~(\ref{Eilenberger_RA2}) and (\ref{Eilenberger_RA3}), Eqs.~(\ref{Eilenberger_K2_g})-(\ref{Eilenberger_K3}), Eqs.~(\ref{gap_equation2}) and (\ref{gap_equation3}), along with the normalization conditions. These equations will be solved in the following sections.

For a semi-infinite superconductor [see Fig.~\ref{fig1} (b)] with a large \(\lambda/\xi\) ratio, 
the aforementioned formulations can be directly applied. 
Given that the current distribution changes gradually over the coherence length, 
the system behaves as locally uniform. 
The primary distinction in this setting is that \(\mathbf{q}\) varies with the distance \(x\) from the surface, reflecting the Meissner current distribution.
In scenarios where the current density is significantly lower than the depairing current density~\cite{2014_Gurevich, 2019_Kubo_Gurevich}, the linear London equation can be employed, yielding \(q = q(0)e^{-x/\lambda}\), where \(\lambda\) represents the London penetration depth. 
Conversely, when the bias DC approaches the depairing current, it becomes necessary to account for the nonlinear Meissner effect to accurately describe the current distribution (see, e.g., Refs.~\cite{Wave_Sauls, 2020_Kubo_2, 2021_Kubo}), but our formulation remains applicable even under these conditions.

\section{Solutions of the Keldysh-Eilenberger equations}

\subsection{Equilibrium Green's functions under a bias DC}

First, we calculate the equilibrium Green's functions under a bias DC without time-dependent perturbation. To determine the equilibrium pair potential for an arbitrary bias, the equilibrium depairing current density \(J_{\rm dp}\), and accurately evaluate the quasiparticle spectrum, we {\it nonperturbatively} solve the equilibrium part of the Keldysh-Eilenberger equations [Eq.~(\ref{Eilenberger_RA2})] under the bias DC~\cite{Jujo, KL, Lin_Gurevich, 2022_Kubo}.

We solve Eq.~(\ref{Eilenberger_RA2}) using the Matsubara formalism:
\begin{eqnarray}
&&\biggl\{ i\frac{\pi (q_b/q_0) \cos \theta}{2} \Delta_0 + \hbar \omega_m  \biggr\} f_m - \Delta_b g_m \nonumber \\
&&= \gamma \bigl( \langle f_m \rangle g_m -  \langle g_m \rangle f_m \bigr).\label{Eilenberger_RA2M}
\end{eqnarray}
Here, \(g_m = g_b(i\hbar\omega_m)\) and \(f_m = i f_b(i\hbar\omega_m)\) represent the normal and anomalous quasiclassical Matsubara Green's functions under a bias DC, respectively. 
The normalization condition is expressed as \(g_m^2 + f_m^2 = 1\), and \(\hbar \omega_m = 2\pi k_B T (m + 1/2)\) defines the Matsubara frequency. 
The parameter \(q_0 = 1/\xi_0 = \pi \Delta_0/\hbar v_f\) denotes the inverse of the BCS coherence length. 
The Eilenberger equation is complemented by the BCS gap equation [Eq.~(\ref{gap_equation2})] or, in the Matsubara formulation,
\begin{equation}
\ln \frac{T_{c}}{T} = 2\pi k_B T \sum_{\omega_m > 0} \left( \frac{1}{\hbar \omega_m} - \frac{\langle f_m \rangle}{\Delta_b} \right).
\label{gap_equation2M}
\end{equation}
Here, \(k_B T_{c} = \Delta_0 \exp(\gamma_E)/\pi \approx \Delta_0/1.76 \) is the critical temperature of a BCS superconductor without a bias DC, where \(\gamma_E = 0.577\) is Euler's constant. 
By solving Eqs.~(\ref{Eilenberger_RA2M}) and (\ref{gap_equation2M}) for the current-carrying state, we can calculate the pair potential under a bias DC, \(\Delta_b\), as a function of the bias superfluid momentum $q_b$.

\begin{figure}[tb]
   \begin{center}
   \includegraphics[width=0.49\linewidth]{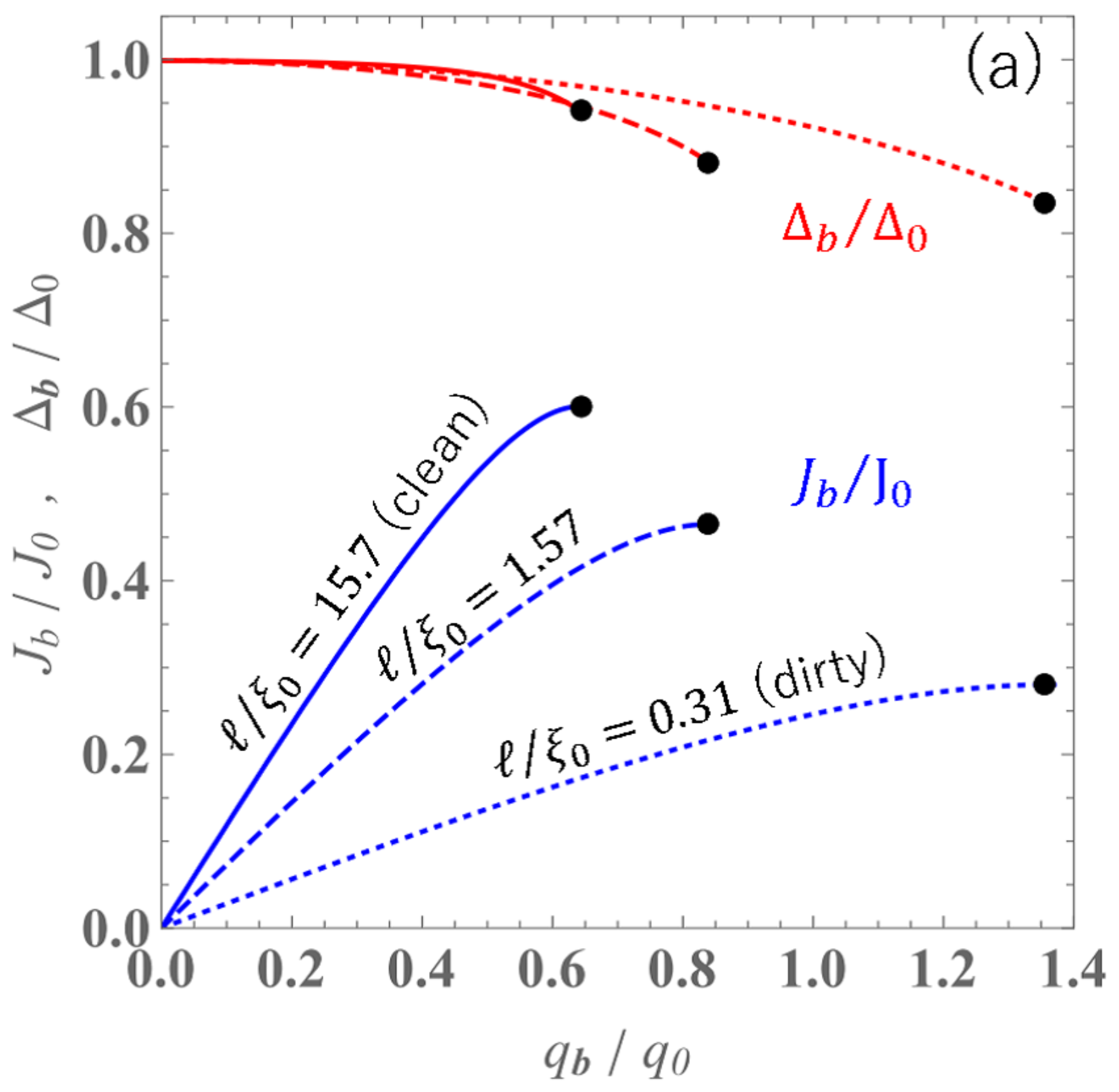}
   \includegraphics[width=0.49\linewidth]{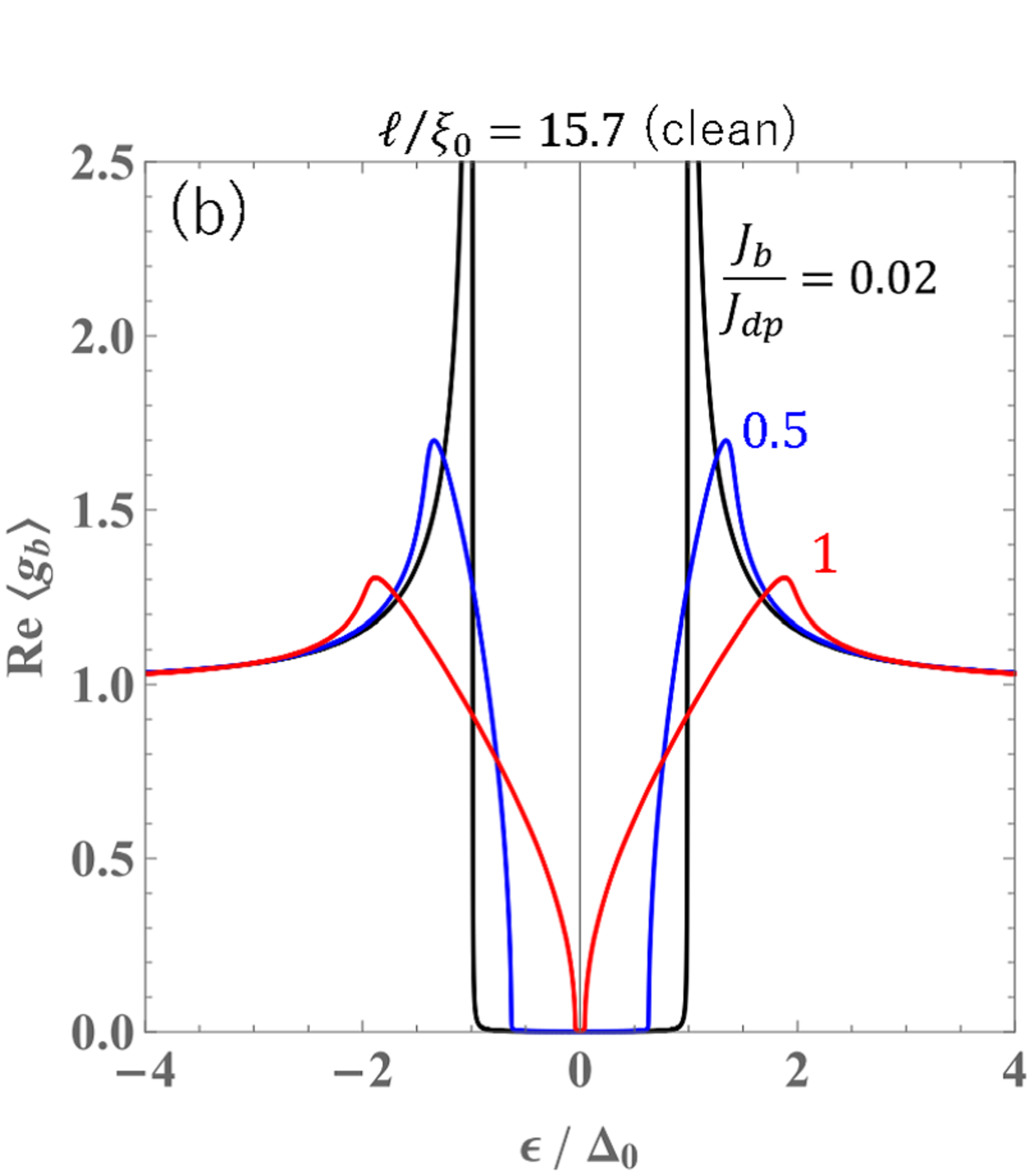}
   \includegraphics[width=0.49\linewidth]{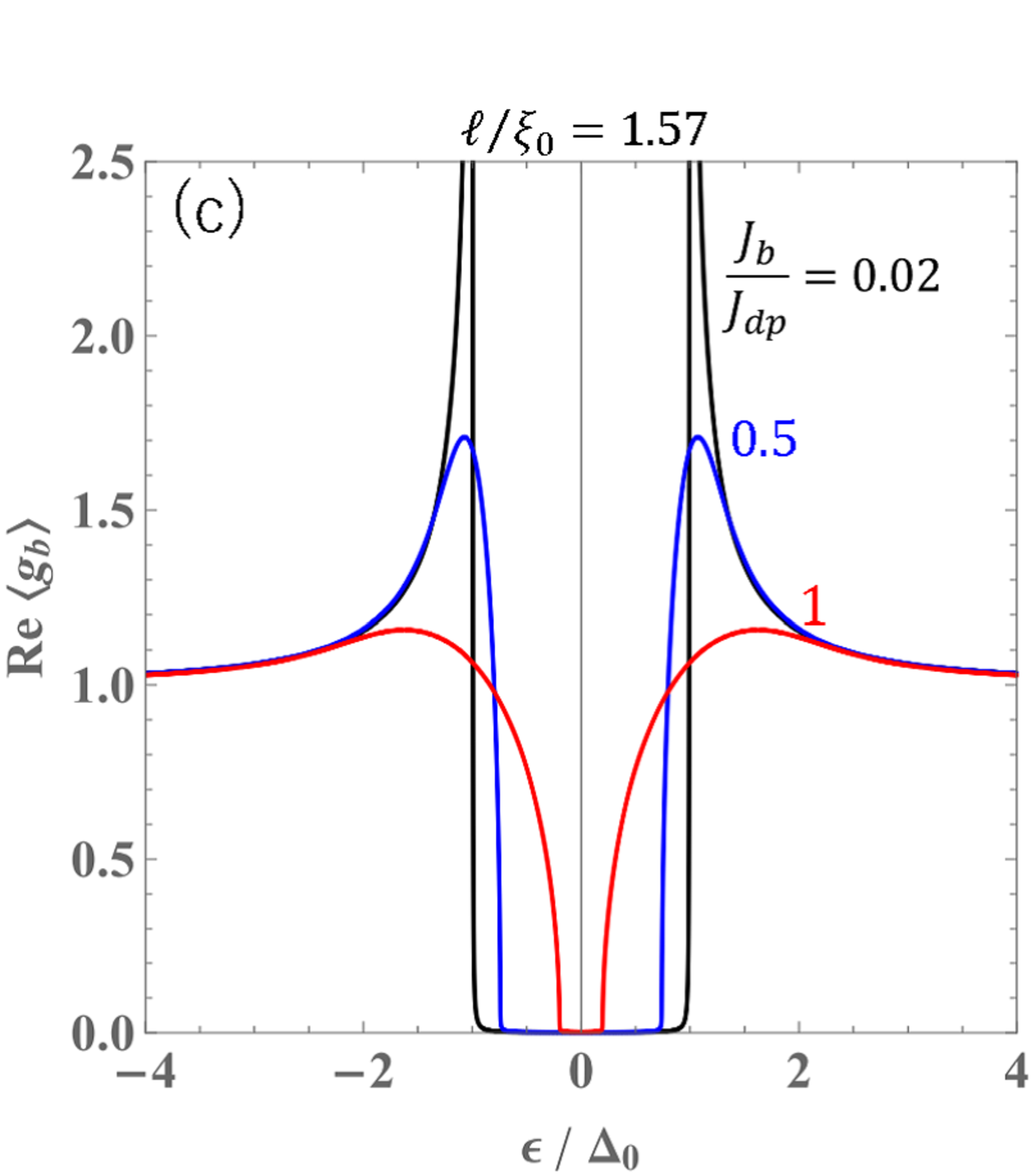}
   \includegraphics[width=0.49\linewidth]{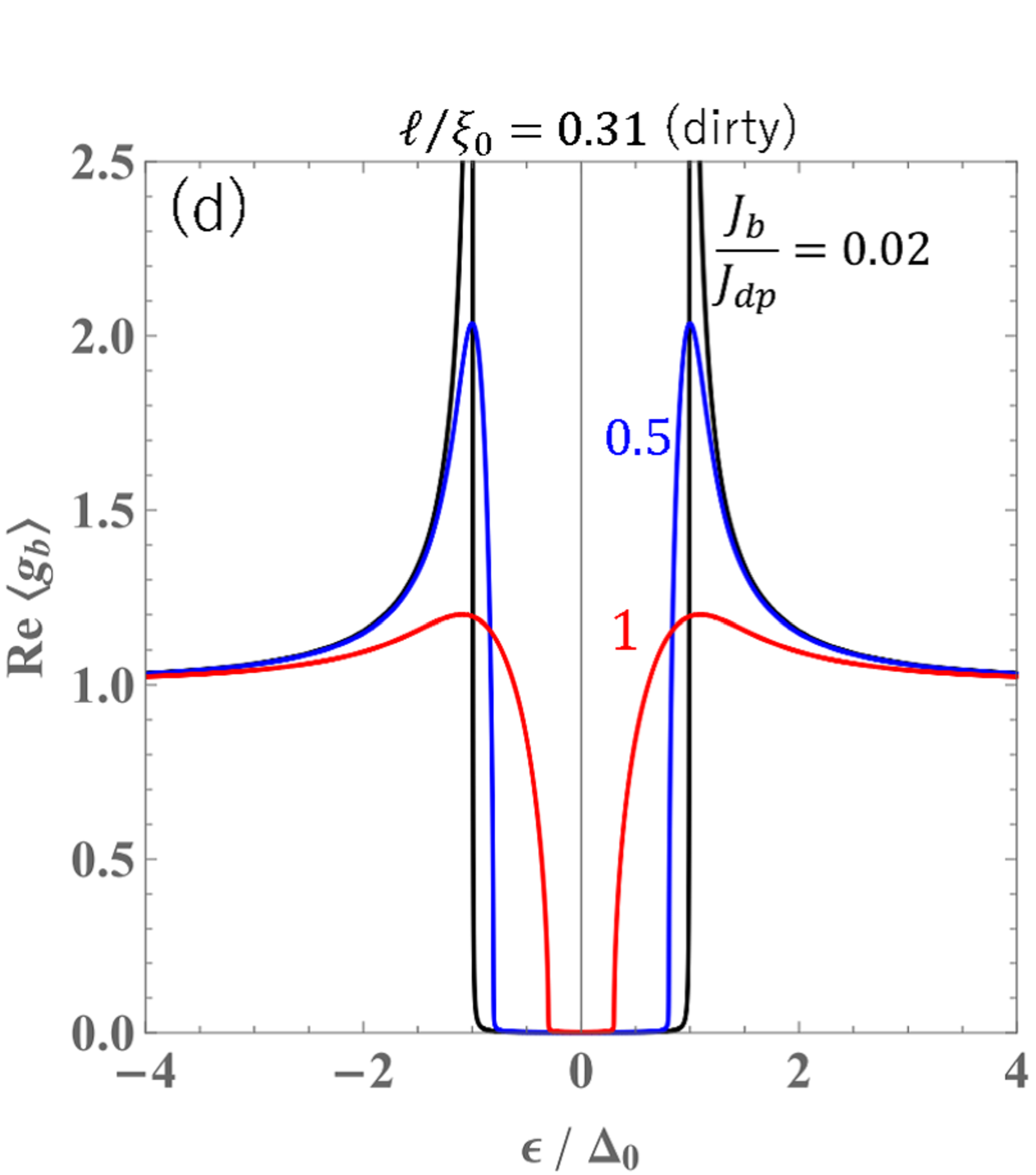}
   \end{center}\vspace{0 cm}
   \caption{
(a) The red and blue curves represent the pair potential \(\Delta_b\) and the current density \(J_b\) as functions of the bias superfluid momentum \(q_b\). The black dots at the tip of the blue curves indicate the equilibrium depairing current density \(J_{\rm dp}\). 
(b-d) The quasiparticle density of states (DOS) for superconductors with mean free paths \(\ell/\xi_0 = 15.7\), \(1.57\), and \(0.31\) under various bias DCs. 
All the calculations in this figure are performed at \(T/T_c = 0.217\). 
   }\label{fig2}
\end{figure}

The solution for Eq.~(\ref{Eilenberger_RA2M}) as given by~Ref.\cite{Lin_Gurevich} is
\begin{eqnarray}
g_m &=& \frac{a_m+i\cos\theta}{\sqrt{(a_m+i\cos\theta)^2 + b_m^2}} ,\\
f_m &=& \frac{b_m}{\sqrt{(a_m+i\cos\theta)^2 + b_m^2}} . 
\end{eqnarray}
Here, the parameters $a_m$, $b_m$, $\langle g_m \rangle$, and $\langle f_m \rangle$ are determined by 
\begin{eqnarray}
&& \frac{\pi \Delta_0}{2}\frac{q_b}{q_0} a_m = \hbar\omega_m  + \langle g_m \rangle \gamma , \label{am}\\
&& \frac{\pi \Delta_0}{2}\frac{q_b}{q_0} b_m = \Delta_b + \langle f_m \rangle \gamma , \label{bm} \\
&& \langle g_m \rangle^4 + (a_m^2 +b_m^2 -1) \langle g_m \rangle^2 -a_m^2=0, \label{gmAve} \\
&& \frac{\langle f_m \rangle}{b_m} = \tan^{-1}\frac{\langle g_m \rangle}{a_m} ,\label{fmAve} 
\end{eqnarray}
where $\Delta_b$ satisfies Eq.~(\ref{gap_equation2M}).

For a given superfluid momentum \( q_b \), the bias current density \( J_b \) can be calculated as ${\bf J}_b = -4\pi k_B T e N_0 {\rm Im} \sum_{\omega_m>0} \langle {\bf v}_{f} g_m \rangle$ (see, e.g., Ref.~\cite{Kopnin}) or equivalently~\cite{2022_Kubo},
\begin{eqnarray}
\frac{J_b}{J_0} = -\frac{\sqrt{6}\pi k_B T}{\Delta_0} \sum_{\omega_m>0}  \int_{-1}^{1} \!\!dc \bigl\{ c \sin u(c) \sin v(c) \bigr\} . \label{DC_current} \nonumber  \\
\end{eqnarray}
Here, $u+iv =\tan^{-1}[b_m/(a_m+ic)]$, 
where \( c = \cos\theta \), 
$J_0=H_{c0}/\lambda_0$, 
and $H_{c0}=\Delta_0\sqrt{N_0/\mu_0}$ represents the thermodynamic critical field of the BCS superconductor in the zero-current state at \( T = 0 \). 
Additionally, \( \lambda_0^{-2} = (2/3) \mu_0 e^2 N_0 v_f^2 \) represents the clean-limit (\( \gamma \to 0 \)) BCS penetration depth at \( T \to 0 \).

Solving Eqs.~(\ref{am})-(\ref{fmAve}) in conjunction with Eq.~(\ref{gap_equation2M}) allows us to obtain the Matsubara Green's functions and the equilibrium pair potential \(\Delta_b(q_b, \ell, T)\) under a bias DC. Substituting this solution into Eq.~(\ref{DC_current}) yields the current density \(J_b(q_b, \ell, T)\).

Fig.~\ref{fig2}(a) displays the equilibrium pair potential \(\Delta_b\) (red curves) and the current density \(J_b\) (blue curves) as functions of the superfluid momentum \(q_b\) for different impurity scattering rates or mean free paths. Note that \(\gamma/\Delta_0 = \pi \xi_0/2\ell\). The maximum value of \(J_b\) is referred to as the depairing current density \(J_{dp}\), indicated by the black dot in the figure. 
For foundational studies on dirty limit superconductors, refer to the seminal works~\cite{1963_Maki} and for superconductors with arbitrary impurity concentrations, see~\cite{KL}. Further detailed analyses of the depairing current density are available in the literature (see, for example, Refs.~\cite{Clem_Kogan, 2020_Kubo_2} for the dirty limit theory and Refs.~\cite{Lin_Gurevich, 2022_Kubo} for the theory applicable to arbitrary impurity concentrations).

The retarded Green's functions, \(g_b\) and \(f_b\), under a bias DC, comply with the real-frequency representation of the Eilenberger equation [Eq.~(\ref{Eilenberger_RA2})]. 
It should be noted that the equilibrium pair potential \(\Delta_b\) under a bias DC has already been calculated as described previously [see Fig.~\ref{fig2} (a)]. The solution to these equations is provided as follows: 
\begin{eqnarray}
g_b  &=& \frac{a-\cos\theta}{\sqrt{(a-\cos\theta)^2-b^2}} , \label{gb}\\
f_b  &=& \frac{b}{\sqrt{(a-\cos\theta)^2-b^2}} . 
\end{eqnarray}
Here, the complex parameters \(a\), \(b\), \(\langle g_b \rangle\), and \(\langle f_b \rangle\) are determined by solving the following set of four complex equations:
\begin{eqnarray}
&&\frac{\pi \Delta_0}{2} \frac{q_b}{q_0} a = \epsilon  + i\gamma \langle g_b \rangle , \label{abgbfb_equations1} \\
&&\frac{\pi \Delta_0}{2} \frac{q_b}{q_0} b= \Delta_b + i\gamma \langle f_b \rangle , \\
&&\langle g_b \rangle = \frac{i}{2} \Bigl[ \sqrt{b^2-(a+1)^2} -\sqrt{b^2-(a-1)^2} \Bigr], \\
&&\biggl[ \biggl( 1- \langle g_b \rangle^2 - \frac{b^2}{2} \biggr) \tanh\frac{2\langle f_b \rangle}{b} + a \langle g_b \rangle \biggr]^2 \nonumber \\
&&= a^2-b^2 +1 -\langle g_b \rangle^2 \label{abgbfb_equations4}. 
\end{eqnarray}

Figures~\ref{fig2}(b)-(d) display the quasiparticle density of states (DOS) under a bias DC, calculated from Eqs.~(\ref{abgbfb_equations1})-(\ref{abgbfb_equations4}) for different mean free paths (\(\gamma \propto \ell^{-1}\)) and varying intensities of bias DC (\(J_b/J_{dp}\)). Each \(J_{dp}\) is calculated as shown in Figure~\ref{fig2}(a). Note that for these numerical calculations (and all subsequent numerical calculations), we introduced a small fixed Dynes parameter \(0.001\Delta_0\) to facilitate the evaluations. These figures illustrate that the evolution of the quasiparticle spectrum under bias DC strongly depends on the mean free path, as previously noted in the literature (see, e.g., Ref.~\cite{Lin_Gurevich}).

We have now obtained the solutions of the equilibrium state Eilenberger equations under an arbitrary strength of bias DC [Eqs.~(\ref{Eilenberger_RA2}), (\ref{Eilenberger_K2_g}), and (\ref{Eilenberger_K2_sigma})], resulting in \(\hat{g}_b^{R, A, K}\) and \(\hat{\sigma}_b^{R, A, K}\). The next step is to solve the Eilenberger equations for the perturbative components as specified in Eqs.~(\ref{Eilenberger_RA3}) and (\ref{Eilenberger_K3}). This will enable us to calculate the complex conductivity of a superconductor with an arbitrary mean free path under a weak to strong bias DC~\cite{Jujo}.

\subsection{Solutions of the Keldysh-Eilenberger equations for the perturbative components} \label{solution_perturbative}

To solve Eqs.~(\ref{Eilenberger_RA3}) and (\ref{Eilenberger_K3}), 
we transition to Fourier space, converting from time domain \(t\) to frequency domain \(\omega\)~\cite{Rainer_Sauls}. 
Utilizing the transformation for the circle product: \((\alpha \circ \beta) (\epsilon, t) \rightarrow \alpha(\epsilon_+) \beta(\epsilon, \omega)\) when \(\alpha = \alpha(\epsilon)\) and \(\beta = \beta(\epsilon, t)\), and \((\alpha \circ \beta) (\epsilon, t) \rightarrow \alpha (\epsilon, \omega) \beta(\epsilon_-)\) when \(\alpha = \alpha (\epsilon, t)\) and \(\beta = \beta(\epsilon)\), where \(\epsilon_{\pm} = \epsilon \pm \omega/2\), we reformulate Eqs.~(\ref{Eilenberger_RA3}) and (\ref{Eilenberger_K3}) as follows:
\begin{eqnarray}
&&\Bigl\{ \bigl( \epsilon_+ - W_b\cos\theta \bigr) \hat{\tau}_3 + \hat{\Delta}_b + i\gamma \langle \hat{g}_b^r(\epsilon_+) \rangle \Bigr\} \delta \hat{g}^r(\epsilon, \omega)  \nonumber \\
&&-\delta \hat{g}^r(\epsilon, \omega) \Bigl\{ \bigl( \epsilon_- - W_b\cos\theta \bigr) \hat{\tau}_3 + \hat{\Delta}_b + i\gamma \langle \hat{g}_b^r(\epsilon_-) \rangle \Bigr\} \nonumber \\
&&+ \hat{g}_b^r(\epsilon_+)  \bigl\{ \delta W(\omega) \cos\theta\hat{\tau}_3 + \delta \hat{\sigma}^r(\epsilon, \omega) \bigr\} \nonumber \\
&&- \bigl\{ \delta W(\omega) \cos\theta \hat{\tau}_3 + \delta \hat{\sigma}^r(\epsilon, \omega) \bigr\} \hat{g}_b^r(\epsilon_-)=0 , 
\end{eqnarray}
and
\begin{eqnarray}
&&\Bigl\{ (\epsilon_+ -W_b\cos\theta) \hat{\tau}_3 - \hat{\sigma}_b^R(\epsilon_+) \Bigr\} \delta \hat{g}^K(\epsilon, \omega)  \nonumber \\
&&- \delta \hat{g}^K(\epsilon, \omega) \Bigl\{ (\epsilon_- -W_b\cos\theta) \hat{\tau}_3 - \hat{\sigma}_b^A(\epsilon_-) \Bigr\}   \nonumber \\
&& + \hat{g}_b^R(\epsilon_+) \delta \hat{\sigma}^K(\epsilon, \omega) 
- \delta \hat{\sigma}^K(\epsilon, \omega) \hat{g}_b^A(\epsilon_-) \nonumber \\
&& + \hat{g}_b^{K}(\epsilon_+) \bigl\{ \delta W(\omega) \cos\theta \hat{\tau}_3 + \delta \hat{\sigma}^A(\epsilon, \omega)  \bigr\}   \nonumber \\
&& - \bigl\{ \delta W(\omega) \cos\theta \hat{\tau}_3 + \delta \hat{\sigma}^R(\epsilon, \omega) \bigr\} \hat{g}_b^{K}(\epsilon_-)  \nonumber \\
&& - \hat{\sigma}_b^K (\epsilon_+) \delta\hat{g}^A(\epsilon, \omega) +\delta\hat{g}^R(\epsilon, \omega) \hat{\sigma}_b^K (\epsilon_-)
=0 .
\end{eqnarray}
The normalization conditions transform as follows: \(\hat{g}_b^r (\epsilon_+) \delta \hat{g}^r(\epsilon, \omega) + \delta \hat{g}^r(\epsilon, \omega) \hat{g}_b^r(\epsilon_-) = 0\) for the retarded and advanced components, and \(\hat{g}_b^R (\epsilon_+) \delta \hat{g}^K(\epsilon, \omega) + \delta \hat{g}^K(\epsilon, \omega) \hat{g}_b^A(\epsilon_-) + \delta \hat{g}^R(\epsilon, \omega) \hat{g}_b^K(\epsilon_-) + \hat{g}_b^K(\epsilon_+) \delta \hat{g}^A (\epsilon, \omega)= 0\) for the Keldysh components. 
Note that the Higgs mode is included in $\delta \sigma = \delta \sigma_{\rm imp} - \delta \Delta$ in the above equations.

After extensive but straightforward calculations, we arrive at the following solutions (see also Ref.~\cite{Jujo}):
\begin{eqnarray}
\delta g^r (\epsilon, \omega) 
&=& \frac{\hat{g}_b^r(\epsilon_+)}{d_+^r + d_-^r} 
\Bigl[  \bigl\{ \delta W(\omega) \cos\theta\hat{\tau}_3 + \delta \hat{\sigma^r}(\epsilon, \omega) \bigr\} \hat{g}_b^r (\epsilon_-) \nonumber \\
&&- \hat{g}_b^r(\epsilon_+) \bigl\{ \delta W(\omega) \cos\theta\hat{\tau}_3 + \delta\hat{\sigma}^r (\epsilon, \omega) \bigr\} \Bigr] , \label{delta gr}
\end{eqnarray}
for the retarded and advanced components, and
\begin{eqnarray}
\delta g^K (\epsilon, \omega) &=& \tanh \frac{\epsilon_-}{2kT} \delta \hat{g}^R(\epsilon, \omega) - \tanh \frac{\epsilon_+}{2kT} \delta \hat{g}^A(\epsilon, \omega) \nonumber \\
&& + \Bigl( \tanh \frac{\epsilon_+}{2kT} - \tanh \frac{\epsilon_-}{2kT} \Bigr) \delta \hat{g}^a(\epsilon, \omega) , \label{delta_g_K} \\
\delta g^a(\epsilon, \omega) &=& \frac{\hat{g}_b^R(\epsilon_+)}{d_+^R + d_-^A} 
\Bigl[ \bigl\{ \delta W(\omega)\cos\theta \hat{\tau}_3 + \delta \hat{\sigma^a}(\epsilon, \omega) \bigr\} \hat{g}_b^A (\epsilon_-) \nonumber \\
&&- \hat{g}_b^R(\epsilon_+) \bigl\{ \delta W(\omega)\cos\theta \hat{\tau}_3 + \delta\hat{\sigma}^a(\epsilon, \omega) \bigr\} \Bigr] ,
\end{eqnarray}
for the Keldysh components. 
Here, 
\begin{eqnarray}
d^R 
&=& \sqrt{(\epsilon - W_b\cos\theta + i\gamma \langle g_{b} \rangle)^2-(\Delta_b + i\gamma \langle f_{b} \rangle)^2} \nonumber \\
&=& (\pi \Delta_0/2)(q_b/q_0) \sqrt{(a-\cos\theta)^2-b^2} = d ,
\end{eqnarray}
$d^A= -d^*$, and $d^{R,A}_{\pm} = d^{R,A}(\epsilon_{\pm})$. 
The corrections to the impurity scattering self-energy are given by
\begin{eqnarray}
\delta \hat{\sigma}_{\rm imp}^{R,A,a} &=& \hat{\tau}_3 X^{R,A,a} + i\hat{\tau}_2 Y^{R,A,a}, \label{delta_sigma_imp} \\
X^R &=& ( i\gamma \chi_3 \Psi + i\gamma \eta_2 + \gamma^2 \chi_1\eta_2 -\gamma^2\chi_3 \eta_3 ) \delta W/D,  \\
Y^R &=& \bigl\{ ( i\gamma \chi_1 -\gamma^2 \chi_1\chi_2 + \gamma^2\chi_3^2)\Psi \nonumber \\
&&+ i\gamma \eta_3 - \gamma^2 \chi_2\eta_3 +\gamma^2\eta_2 \chi_3 )\bigr\} \delta W/D , \\
X^a &=& ( i\gamma \chi_3^a \Psi + i\gamma \eta_2^a + \gamma^2 \chi_1^a\eta_2^a -\gamma^2\chi_3^a \eta_3^a ) \delta W/D^a, \nonumber \\
Y^a &=& \bigl\{ ( i\gamma \chi^a_1 -\gamma^2 \chi_1^a \chi_2^a + \gamma^2\chi_3^{a2})\Psi \nonumber \\
&&+ i\gamma \eta_3^a - \gamma^2 \chi_2^a \eta_3^a +\gamma^2\eta_2^a \chi_3^a )\bigr\} \delta W/D^a ,
\end{eqnarray}
$X^A=X^{R*}$, $Y^A=Y^{R*}$,
and
\begin{eqnarray}
(\chi_1, \eta_1)
&=& \Bigl\langle (1, \cos\theta) \, \frac{g_+ g_- + f_+ f_- +1}{d_+ + d_-} \Bigr\rangle, \\
(\chi_2, \eta_2) &=& \Bigl\langle (1, \cos\theta) \,  \frac{g_+ g_- + f_+ f_- -1}{d_+ + d_-} \Bigr\rangle,\\
(\chi_3, \eta_3) &=& \Bigl\langle (1, \cos\theta) \,  \frac{g_+ f_- + f_+ g_-}{d_+ + d_-} \Bigr\rangle, \\
(\chi_1^a, \eta_1^a) &=& \Bigl\langle (1, \cos\theta) \,  \frac{-g_+ g_-^* - f_+ f_-^* +1}{d_+ - d_-^*} \Bigr\rangle, \\
(\chi_2^a, \eta_2^a) &=& \Bigl\langle (1, \cos\theta) \,  \frac{-g_+ g_-^* - f_+ f_-^* -1}{d_+ - d_-^*} \Bigr\rangle ,\\
(\chi_3^a, \eta_3^a) &=& \Bigl\langle (1, \cos\theta) \,  \frac{-g_+ f_-^* - f_+ g_-^*}{d_+ - d_-^*} \Bigr\rangle, \\
D &=& \gamma^2 \chi_3^{2} - (1-i\gamma\chi_1)(1+i\gamma\chi_2), \\
D^a&=& \gamma^2 \chi_3^{a2} - (1-i\gamma\chi_1^a)(1+i\gamma\chi_2^a), 
\end{eqnarray}
The Higgs mode in the frequency domain is given by
\begin{eqnarray}
\delta \Delta (\omega) &=& \Psi \delta W(\omega),  \\
\Psi &=& \frac{(g/4) \int d\epsilon \psi_N(\epsilon)}{1-(g/4)\int d\epsilon \psi_D(\epsilon)}, \label{Psi}\\
\psi_N &=& \kappa_{(\Psi)} \tanh \frac{\epsilon_-}{2kT}  + \kappa_{(\Psi)}^* \tanh \frac{\epsilon_+}{2kT}  \nonumber \\
&& + \Bigl( \tanh \frac{\epsilon_+}{2kT} - \tanh \frac{\epsilon_-}{2kT} \Bigr)\kappa_{(\Psi)}^a , \\
\psi_D &=& \zeta \tanh \frac{\epsilon_-}{2kT}  + \zeta^* \tanh \frac{\epsilon_+}{2kT}  \nonumber \\
&& + \Bigl( \tanh \frac{\epsilon_+}{2kT} - \tanh \frac{\epsilon_-}{2kT} \Bigr)\zeta^a , 
\end{eqnarray}
Here, 
\begin{eqnarray}
&&\kappa_{(\Psi)} = \bigl\{ -\eta_3 + i\gamma (\eta_2 \chi_3 -\chi_2 \eta_3) \bigr\} / D , \label{kappa_psi}\\
&&\kappa_{(\Psi)}^a = \bigl\{ -\eta_3^a + i\gamma (\eta_2^a \chi_3^a -\chi_2^a \eta_3^a) \bigr\} / D^a , \label{kappa_psi_a}\\
&&\zeta = \bigl\{ -\chi_1 + i\gamma (\chi_3^2 -\chi_1 \chi_2) \bigr\} / D , \\
&&\zeta^a = \bigl\{ -\chi_1^a + i\gamma (\chi_3^{a2} -\chi_1^a \chi_2^a) \bigr\} / D^a . \label{zetaa}
\end{eqnarray}
Note the BCS coupling constant $g$ can be eliminated from Eq.~(\ref{Psi}) using the zero-current gap equation $1/g=\int_{\Delta}^{\infty}\tanh(\epsilon/2kT)/\sqrt{\epsilon^2 - \Delta^2}d\epsilon$.

Before concluding this section, let us consider the zero-bias case (\(q_b = 0\)). In this scenario, the equilibrium Green's functions lose their angular dependencies, resulting in \(\eta_{1,2,3} \propto \langle \cos\theta \rangle = 0\) and \(\eta_{1,2,3}^a \propto \langle \cos\theta \rangle = 0\). Consequently, from Eq.~(\ref{Psi}), we find that \(\Psi = 0\) and \(\delta \Delta = 0\): the Higgs mode loses its linear coupling to the electromagnetic perturbation. For \(\eta_{1,2,3} = \eta_{1,2,3}^a = \Psi = 0\), Eq.~(\ref{delta_sigma_imp}) implies that the corrections to the impurity scattering self-energy vanish (\(\delta \sigma_{\rm imp}^{R,A,K} = 0\)). 
Conversely, when a bias DC is applied, we cannot justify disregarding the Higgs mode \(\delta \Delta\) and the impurity scattering self-energy corrections \(\delta \sigma_{\rm imp}^{R, A, K}\).

We have now obtained the solutions of the time-dependent perturbative components of the Eilenberger equations, given by Eqs.~(\ref{delta gr})-(\ref{zetaa}), including the impurity scattering self-energy corrections \(\delta \sigma_{\rm imp}^{R, A, K}\) and the Higgs mode \(\delta \Delta\). In the next section, we will use these results to calculate the complex conductivity under a bias DC.

\section{Complex conductivity formula under a bias DC and time-dependent perturbation}

\subsection{Complex conductivity formula}

The current density coming from the time-dependent perturbation $\delta W$ can be calculated from the formula (see, e.g., Refs.~\cite{Kopnin, Rainer_Sauls})
\begin{eqnarray}
{\bf J}(\omega)
&=& -\frac{eN_0}{4} \int d\epsilon \Bigl\langle {\bf v}_f {\rm Tr}\bigl\{ \hat{\tau}_3 \delta \hat{g}^K(\epsilon,\omega) \bigr\} \Bigr\rangle .
\label{noneqJ}
\end{eqnarray}
Substituting Eq.~(\ref{delta_g_K}) into Eq.~(\ref{noneqJ}) and using $\delta W =(\hbar/2)v_f \delta q = (iev_f/\omega) E$, where $E$ is the electric field, we get the complex conductivity $\sigma=\sigma_1 + i\sigma_2$ as (see also the surface resistance calculation in Ref.~\cite{Jujo})
\begin{eqnarray}
&&\sigma = \sigma^{(0)} + \sigma^{\rm (imp)} + \Psi \sigma^{(\Psi)}, \\
&&\sigma^{(i)} = \frac{-3i \sigma_n}{4\omega \tau} \int d\epsilon \Bigl\{ \kappa_{(i)} \tanh \frac{\epsilon_-}{2kT}  + \kappa_{(i)}^* \tanh \frac{\epsilon_+}{2kT}  \nonumber \\
&& + \Bigl( \tanh \frac{\epsilon_+}{2kT} - \tanh \frac{\epsilon_-}{2kT} \Bigr)\kappa_{(i)}^a \Bigr\}  \hspace{0.7cm} (i=0, {\rm imp}, \Psi). \nonumber \\
\end{eqnarray}
Here, 
\begin{eqnarray}
&&\kappa_{(0)} = \Bigl\langle  c^2 \frac{g_+ g_- + f_+ f_- -1}{d_+ + d_-} \Bigr\rangle,\\
&&\kappa_{(0)}^a = \Bigl\langle  c^2 \frac{-g_+ g_-^* - f_+ f_-^+ -1}{d_+ - d_-^*} \Bigr\rangle,\\
&&\kappa_{\rm (imp)} = \frac{i\gamma}{D} \bigl\{ \eta_2^2 -\eta_3^2 -i\gamma (\chi_1 \eta_2^2 - 2\eta_2 \chi_3 \eta_3 + \chi_2 \eta_3^2) \bigr\} ,  \nonumber \\ \label{kappa_imp} \\
&&\kappa_{\rm (imp)}^a = \frac{i\gamma}{D^a} \bigl\{ \eta_2^{a2} -\eta_3^{a2}  \nonumber  \\ 
&&-i\gamma (\chi_1^a \eta_2^{a2} - 2\eta_2^a \chi_3^a \eta_3^a + \chi_2^a \eta_3^{a2}) \bigr\} , \label{kappa_impa}  
\end{eqnarray}
and $\kappa_{(\Psi)}$ and $\kappa_{(\Psi)}^a$ are given by Eqs.~(\ref{kappa_psi}) and (\ref{kappa_psi_a}). 
The parameters necessary to calculate the complex conductivity are summarized in Table~1.

\begin{table}[t]
\centering
\begin{tabular}{|c|c|}
\hline
\textbf{Parameter} & \textbf{Description} \\
\hline
$T$       & Temperature \\
\hline
$\omega$  & Frequency \\
\hline
$\ell$ (or $\gamma$)    & Mean free path ($\gamma/\Delta_0 = \pi \xi_0/2\ell$) \\
\hline
$J_b$ (or $q_b$) & bias current (or bias superfluid momentum) \\
\hline
\end{tabular}
\caption{Parameters used in the complex conductivity formula.}
\end{table}

Before proceeding to concrete calculations, let us consider the zero-bias case (\(q_b = 0\)). As seen in Section~\ref{solution_perturbative}, in this scenario, the equilibrium Green's functions lose their angular dependencies, causing \(\eta_{1,2,3}\) and \(\eta_{1,2,3}^a\) to vanish. Additionally, \(\kappa_{(0)}\) and \(\kappa_{(0)}^a\) simplify to \(\kappa_{(0)} = (1/3) (g_+ g_- + f_+ f_- - 1) / (d_+ + d_-)\) and \(\kappa_{(0)}^a = (1/3) (-g_+ g_-^* - f_+ f_-^* - 1) / (d_+ - d_-^*)\), respectively. 
Then, from Eqs.~(\ref{kappa_psi}) and (\ref{kappa_psi_a}), and Eqs.~(\ref{kappa_imp}) and (\ref{kappa_impa}), we find that \(\kappa_{\rm (imp)} = \kappa_{\rm (imp)}^a = \kappa_{(\Psi)} = \kappa_{(\Psi)}^a = 0\), leading to \(\sigma = \sigma^{(0)}\), which reproduces the well-known complex conductivity formula for a superconductor with an arbitrary mean free path in the zero-bias case~\cite{Rainer_Sauls, Zimmermann, Herman, Ueki, 2022_Kubo}.

\begin{figure}[tb]
   \begin{center}
   \includegraphics[width=0.49\linewidth]{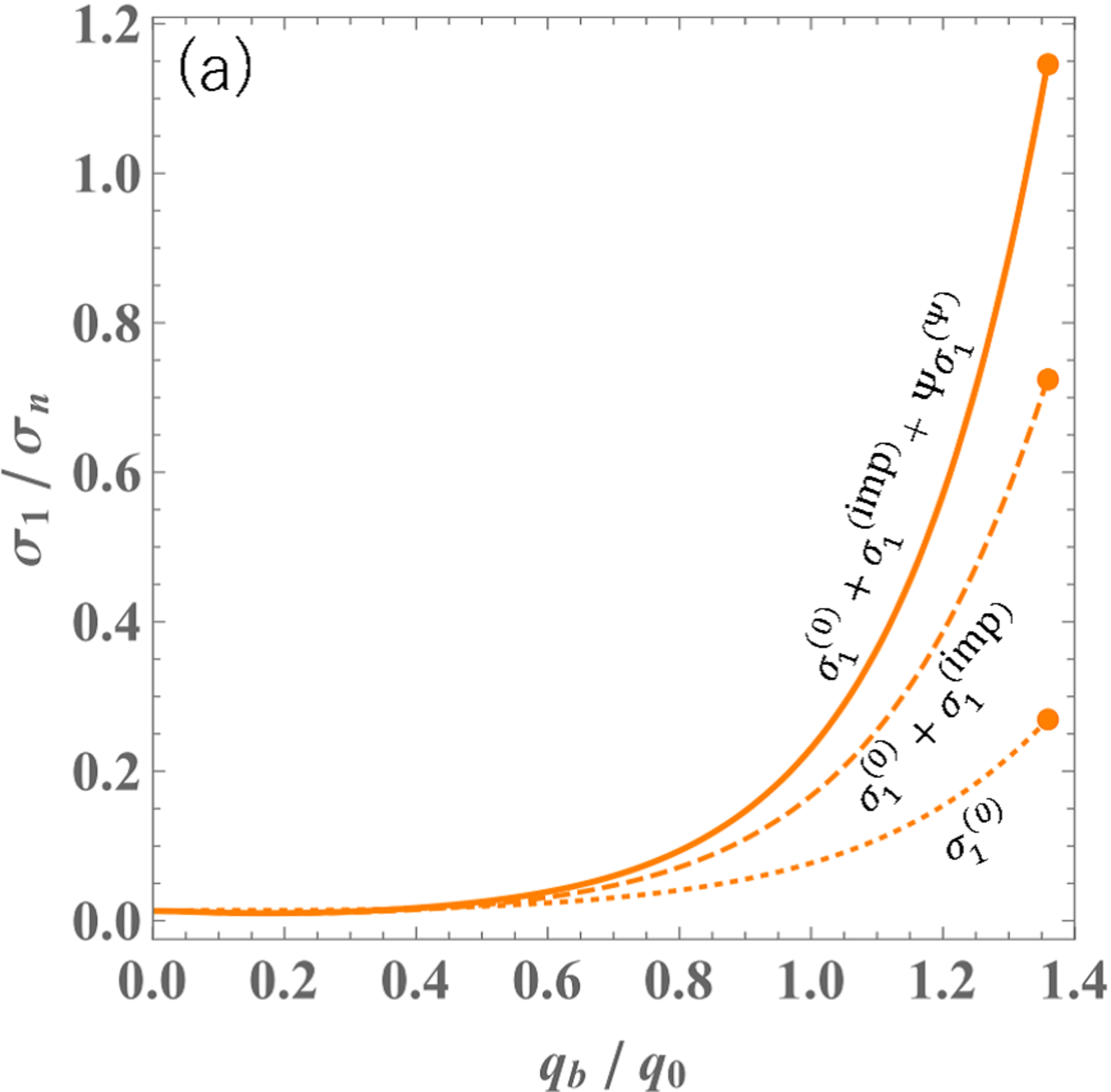}
   \includegraphics[width=0.49\linewidth]{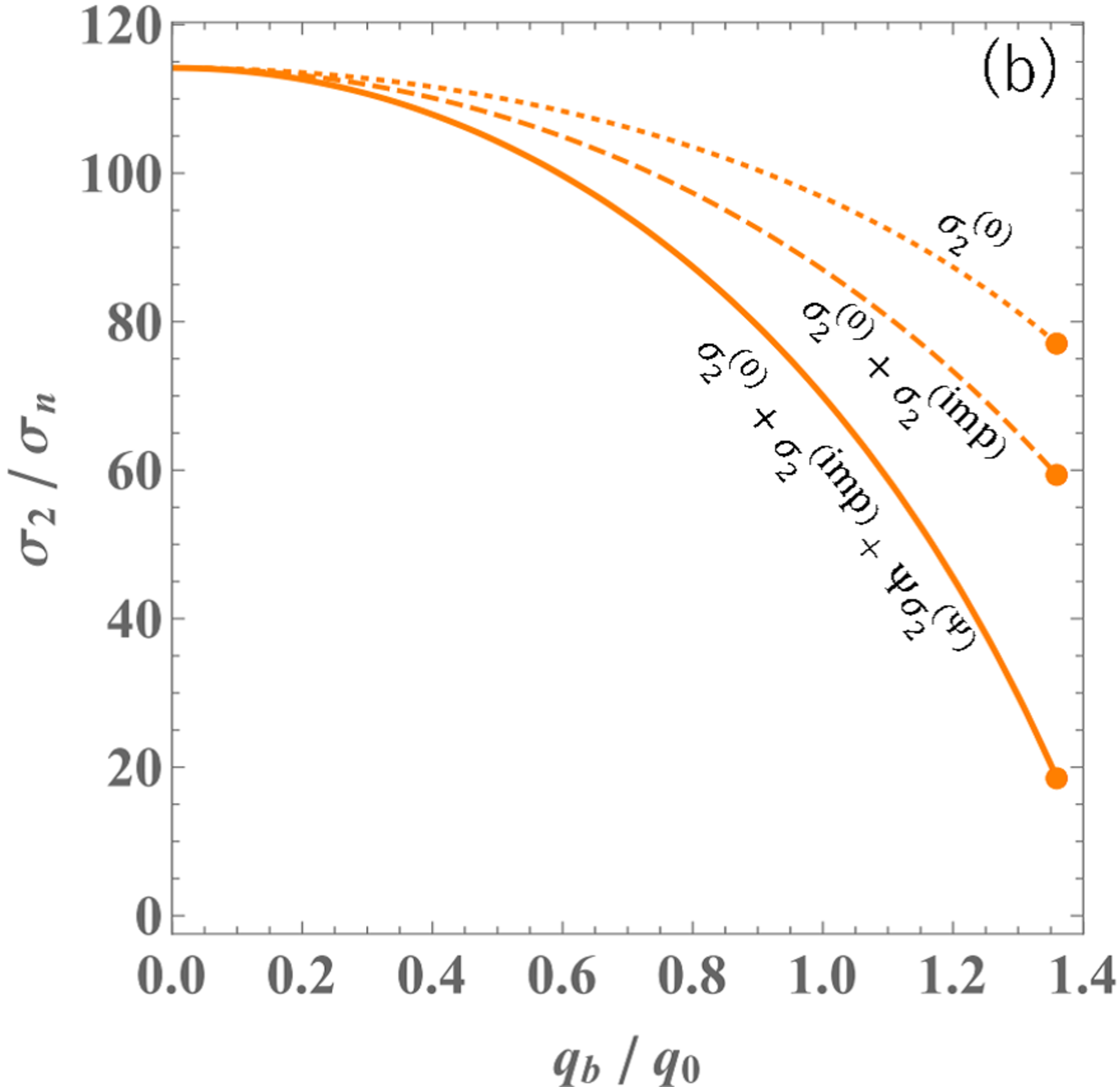}
   \end{center}\vspace{0 cm}
   \caption{
(a) Real part and (b) imaginary part of the complex conductivity in a moderately dirty superconductor ($\ell/\xi_0=0.31$) as functions of bias superfluid momentum,
calculated for $\hbar\omega/\Delta_0 =0.02$ and $T/T_c=0.217$. 
The dots at the edge represent the equilibrium depairing momentum calculated in Fig.~\ref{fig2}(a). 
The solid curves represent the correct calculations \(\sigma_{1,2} = \sigma_{1,2}^{(0)} + \sigma_{1,2}^{\rm (imp)} + \Psi \sigma_{1,2}^{(\Psi)}\), 
while the dashed and dotted curves represent incomplete calculations, omitting one or both of \(\sigma_{1,2}^{\rm (imp)}\) and \(\sigma_{1,2}^{(\Psi)}\). 
   }\label{fig3}
\end{figure}

For a finite bias DC (\(q_b > 0\)), it is not justified to neglect the contributions from \(\sigma^{\rm (imp)}\) and \(\sigma^{(\Psi)}\). In fact, as shown by Moor for the dirty limit~\cite{Moor} and by Jujo for an arbitrary impurity concentration~\cite{Jujo}, these contributions are significant. In particular, the Higgs mode contribution \(\sigma^{(\Psi)}\) results in a characteristic peak in \(\sigma_1\) and a dip in \(\sigma_2\) around the resonant frequency of the Higgs mode \(\hbar \omega/\Delta_0 \simeq 2\), which were later observed experimentally~\cite{Nakamura}. 
This significance extends to lower frequencies (\(\ll \Delta\)) in both the $\sigma_1$ and $\sigma_2$ calculations, as clearly shown in Figure~\ref{fig3}.

In the following sections, we will not only reproduce the peaks and dips around the resonance frequency of the Higgs mode but also investigate the detailed low-frequency behaviors of \(\sigma_{1,2}\), including contributions from both \(\sigma^{\rm (imp)}\) and \(\sigma^{(\Psi)}\).

\subsection{Real part of complex conductivity: $\sigma_1$}\label{section_sigma1}

\begin{figure}[tb]
   \begin{center}
   \includegraphics[width=0.49\linewidth]{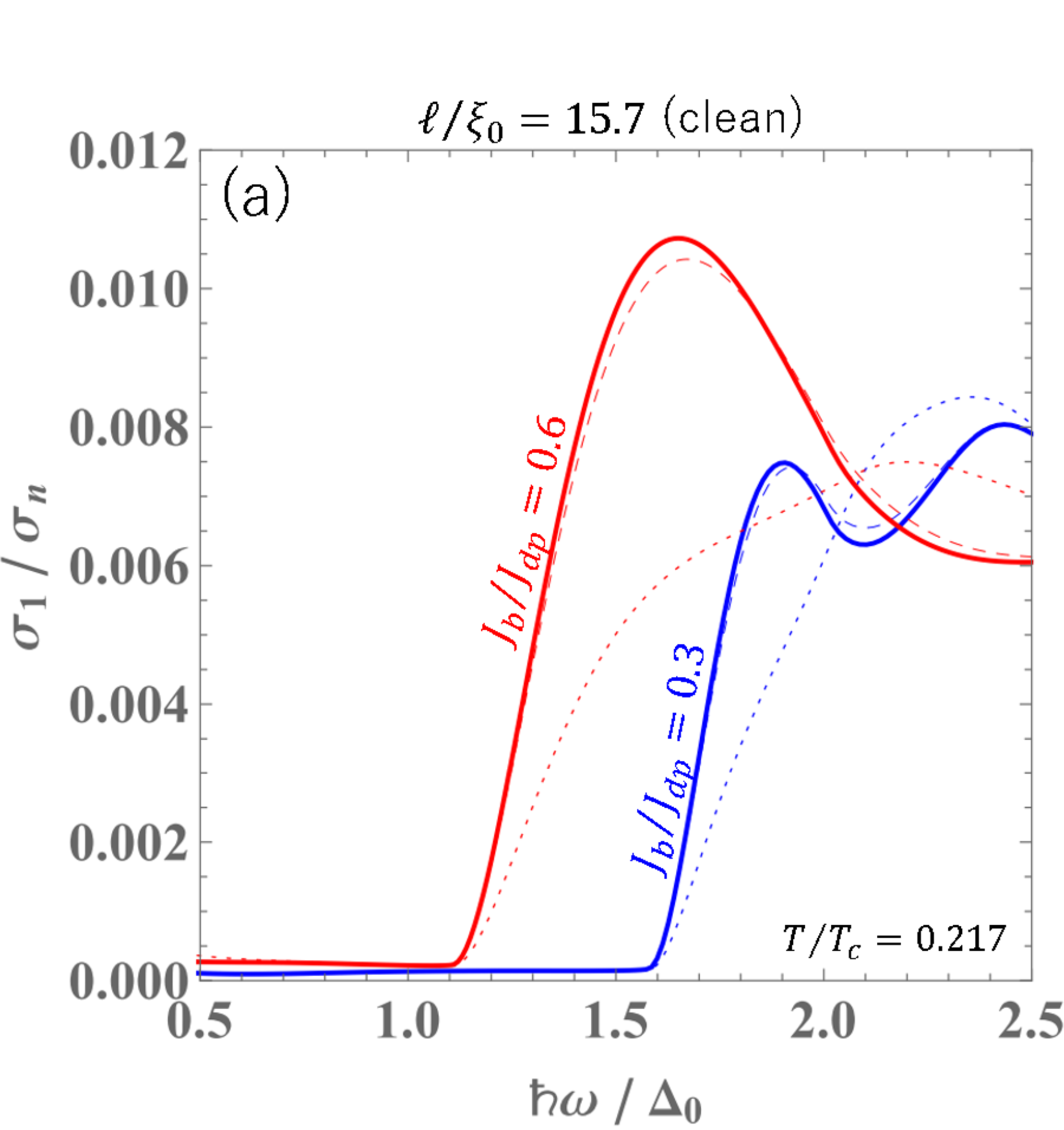}
   \includegraphics[width=0.49\linewidth]{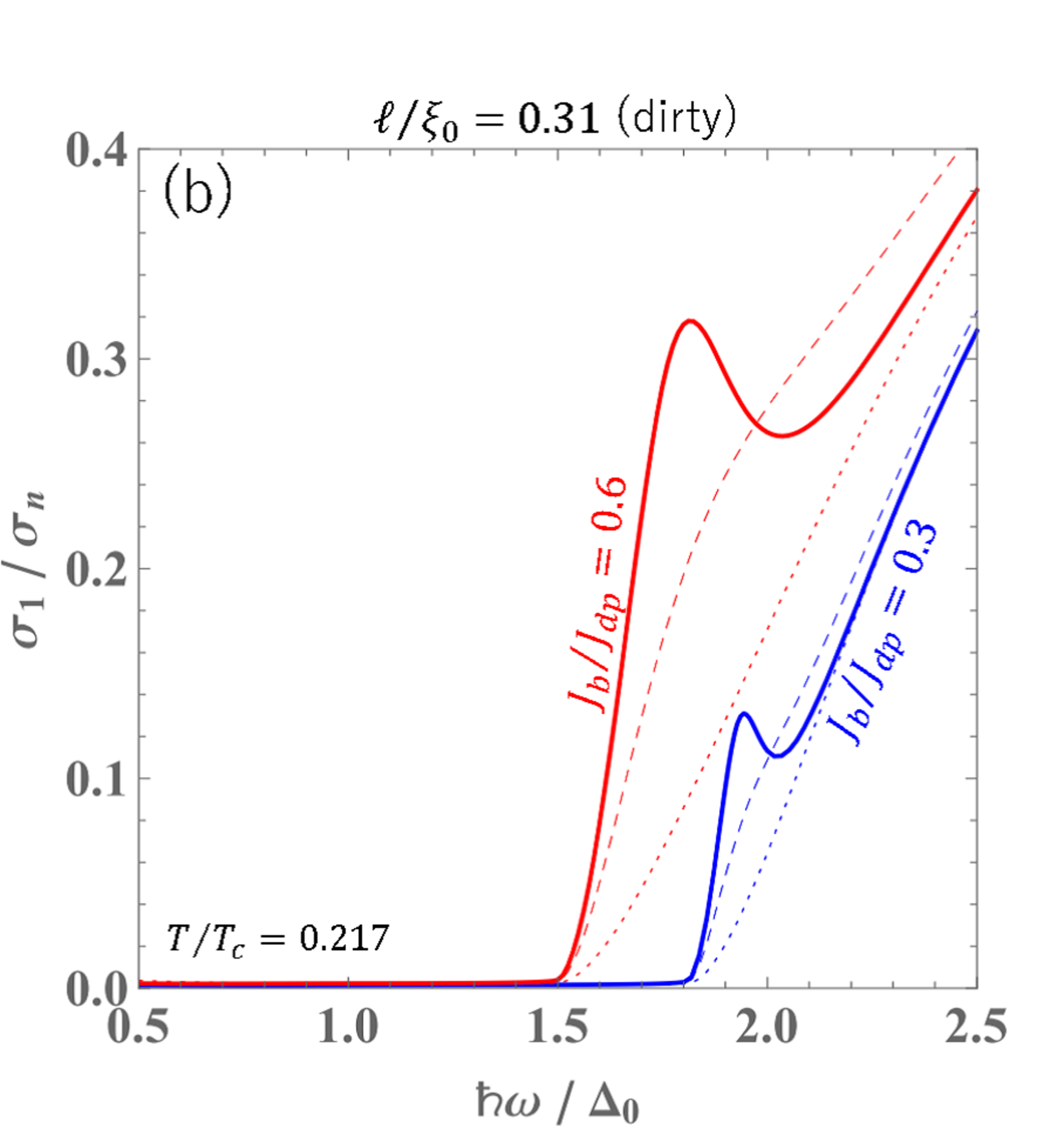}
   \end{center}\vspace{0 cm}
   \caption{
Frequency scan of the real part of the complex conductivity, \(\sigma_1/\sigma_n\), for (a) a clean and (b) a dirty superconductor at different intensities of bias DC, \(J_b/J_{\rm dp}\). 
Here, \(J_{\rm dp}\) is the equilibrium depairing current density calculated in Fig.~\ref{fig2}. 
The dotted, dashed, and solid curves represent \(\sigma_{1,2}^{(0)}\), \(\sigma_{1,2}^{(0)} + \sigma_{1,2}^{\rm (imp)}\), and \(\sigma_{1,2} = \sigma_{1,2}^{(0)} + \sigma_{1,2}^{\rm (imp)} + \Psi \sigma_{1,2}^{(\Psi)}\), respectively. 
The characteristic peak attributed to the Higgs mode is clearly visible in the dirty superconductor. For more details, see the previous study~\cite{Jujo}.
   }\label{fig4}
\end{figure}

Now, let us explore the real part of the complex conductivity, \(\sigma_1\), focusing initially on its frequency dependence. Figure~\ref{fig4} shows \(\sigma_1\) as a function of \(\omega\), replicating the results obtained in the previous study~\cite{Jujo}. For both clean and dirty superconductors, the results that include the contributions from the Higgs mode and self-energy corrections (solid curves) exhibit significant differences compared to those that omit these contributions (dotted curves).
In the clean case, the Higgs mode contribution is minimal. 
This phenomenon is attributed to the Galilean invariance of clean superconductors: switching to a moving frame eliminates the condensate velocity induced by the external field, leaving the order parameter amplitude unchanged, as discussed in the study of the nonlinear electromagnetic response~\cite{Silaev}.
Conversely, in the dirty case, where Galilean invariance is broken, a characteristic peak emerges due to the Higgs mode contribution \(\sigma_1^{(\Psi)}\) at the resonant frequency of the Higgs mode (\(\hbar \omega/\Delta_0 \simeq 2\)), as indicated by the solid curve compared to the dashed curve (see also Ref.~\cite{Nakamura} for the experimental observation).

\begin{figure}[tb]
   \begin{center}
   \includegraphics[height=0.53\linewidth]{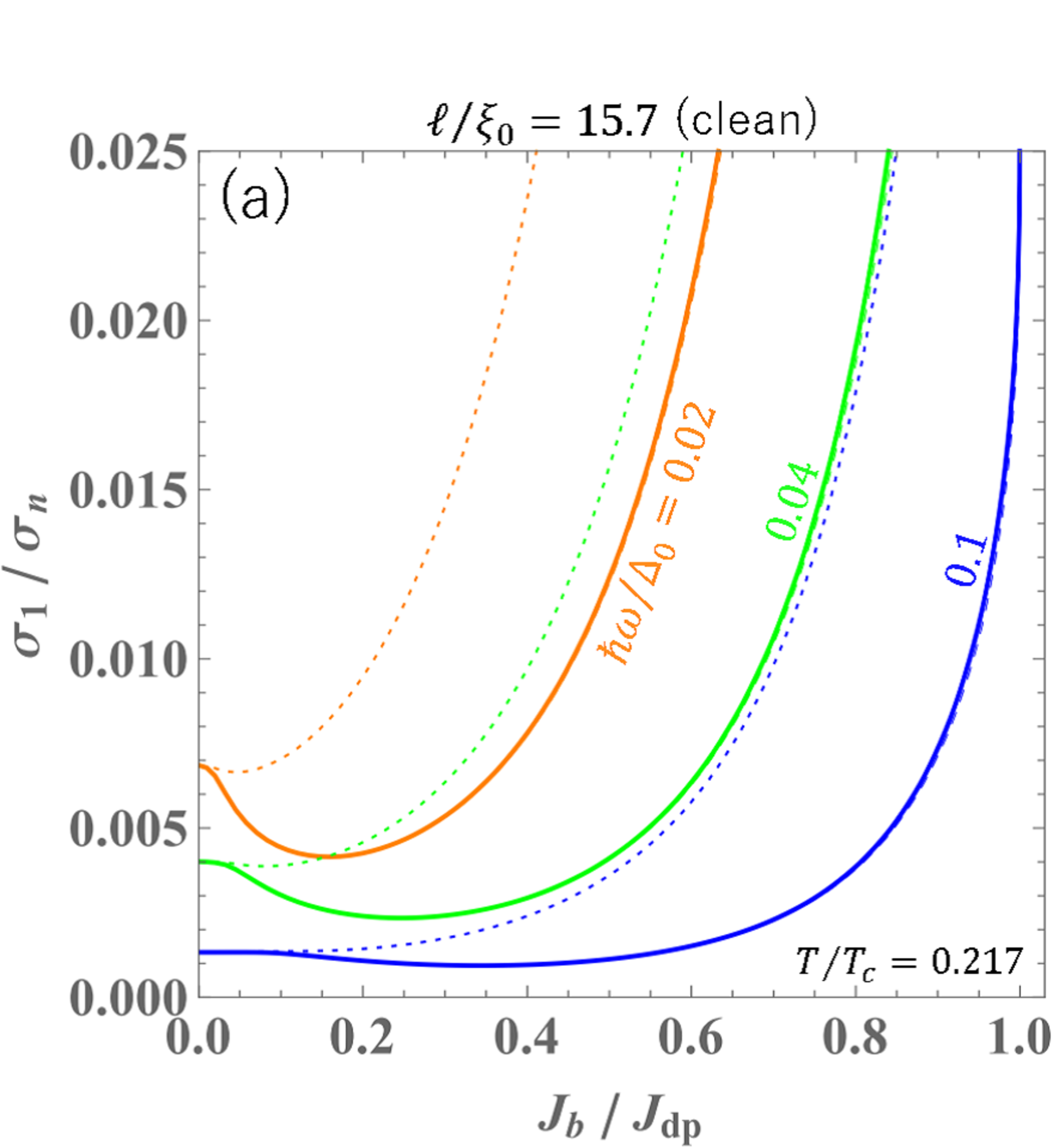}
   \includegraphics[height=0.53\linewidth]{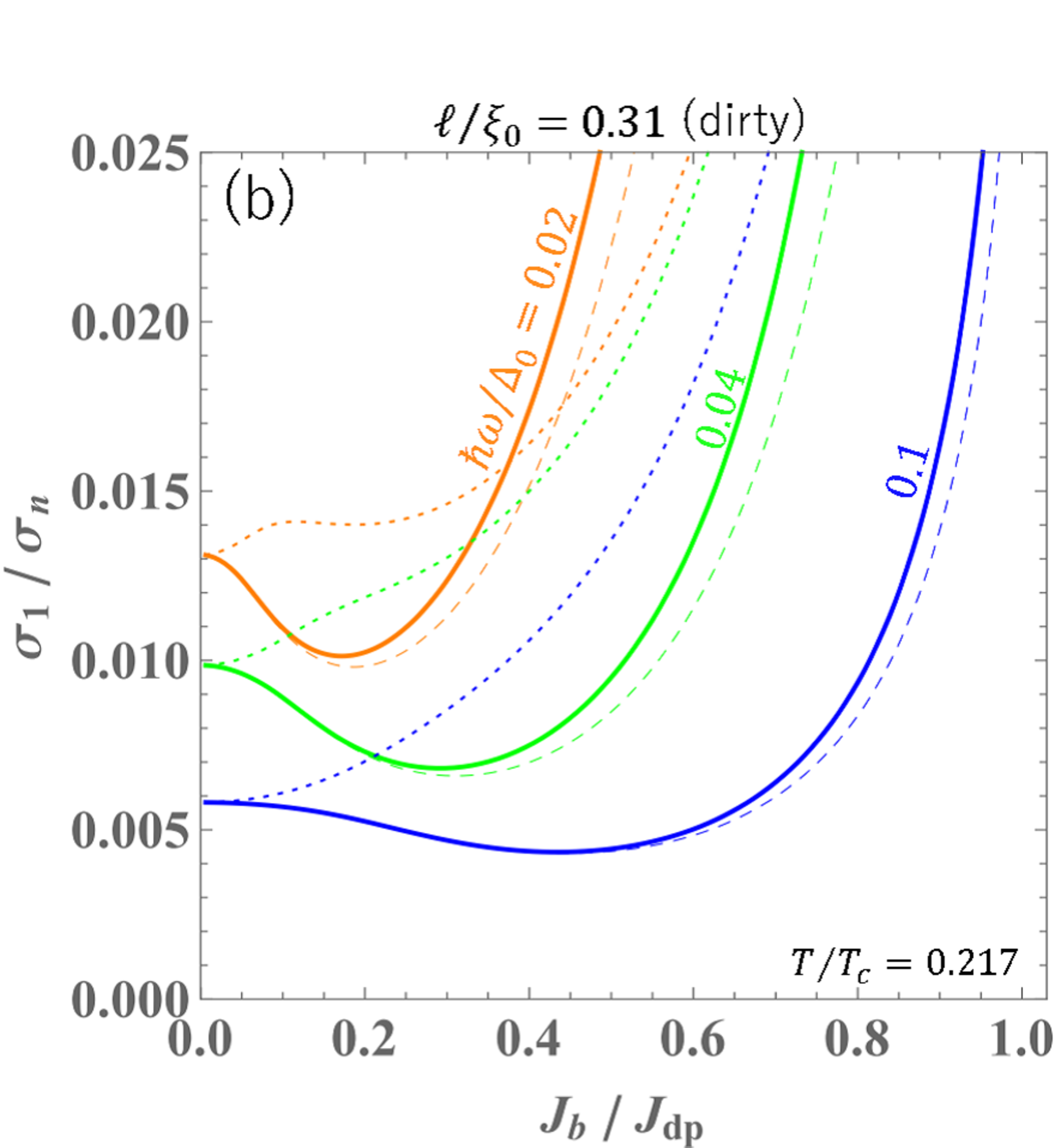}
   \includegraphics[height=0.53\linewidth]{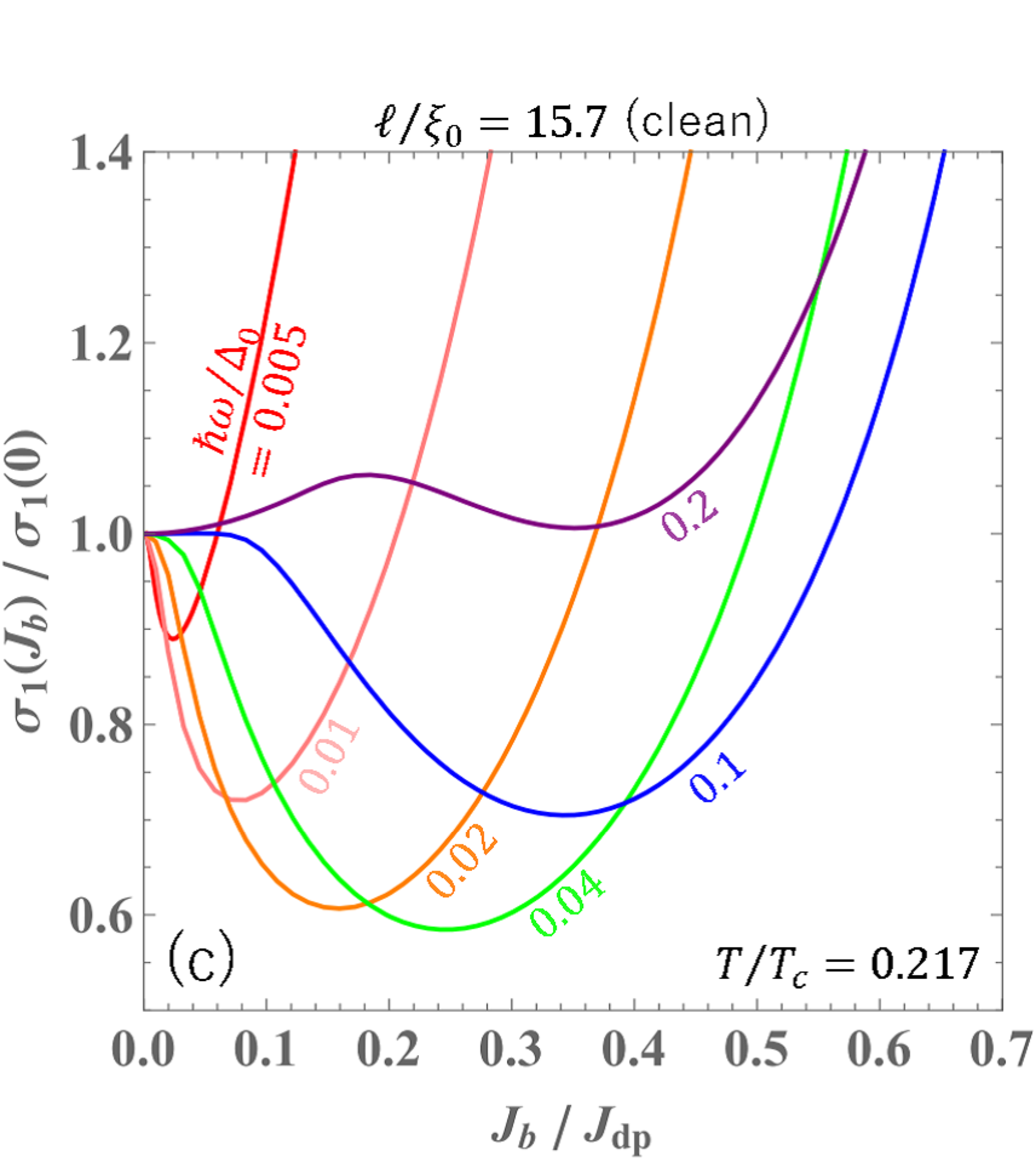}
   \includegraphics[height=0.53\linewidth]{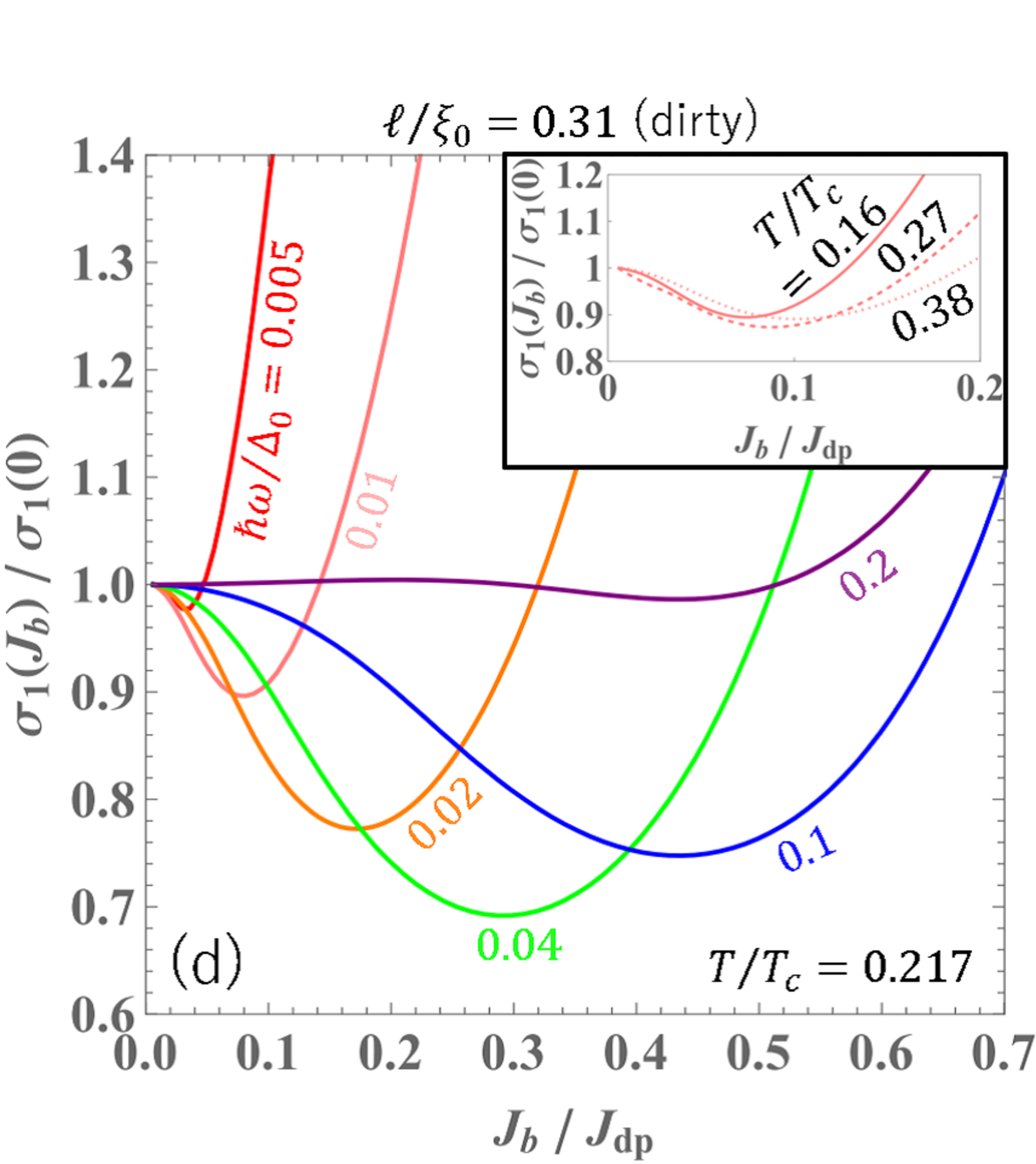}
   \end{center}\vspace{0 cm}
   \caption{
The top panels show the real part of the complex conductivity as functions of bias DC, \(J_b/J_{\rm dp}\), for (a) a clean and (b) a dirty superconductor. The dotted, dashed, and solid curves represent \(\sigma_{1}^{(0)}\), \(\sigma_{1}^{(0)} + \sigma_{1}^{\rm (imp)}\), and \(\sigma_{1} = \sigma_{1}^{(0)} + \sigma_{1}^{\rm (imp)} + \Psi \sigma_{1}^{(\Psi)}\), respectively. The bottom panels display the real part of the complex conductivity normalized by zero bias values, \(\sigma_1(J_b)/\sigma_1(0)\), as functions of bias DC, \(J_b/J_{\rm dp}\), for (c) a clean and (d) a dirty superconductor. This format facilitates the comparison of the current-dependent \(\sigma_1\) reduction rates (dip depths) across different frequencies. Here, \(J_{\rm dp}\) is the equilibrium depairing current density calculated in Fig.~\ref{fig2}. 
The inset shows the results of $\hbar\omega/\Delta_0 = 0.01$ for different temperatures. 
   }\label{fig5}
\end{figure}

Figures~\ref{fig5}(a) and (b) illustrate the bias DC dependence of \(\sigma_1\) at lower frequencies, demonstrating that impurity scattering self-energy corrections are significant not only at high frequencies near the resonant frequency of the Higgs mode (\(\hbar \omega/\Delta_0 \simeq 2\)) but also at lower frequencies (\(\hbar \omega/\Delta_0 \ll 1\)). In the dirty case, the Higgs mode contribution affects the dissipative conductivity, as evidenced by the differences between the solid and dashed curves. Previous studies that did not consider the Higgs mode and self-energy corrections observed that as \(\omega\) increases, the reduction rate decreases, with the dip disappearing at \(\hbar \omega \sim 0.01\) in this temperature regions~\cite{2014_Gurevich, 2020_Kubo_1}. However, as demonstrated in Figures~\ref{fig5}(a) and (b), including these corrections [\(\sigma_1^{\rm (imp)}\) and \(\sigma_1^{(\Psi)}\)] significantly alters this behavior.

Figures~\ref{fig5}(c) and (d) show the real part of the complex conductivity normalized by zero bias values, \(\sigma_1(J_b)/\sigma_1(0)\), as functions of bias DC. The current-dependent reduction of \(\sigma_1\) continues up to \(\hbar \omega \sim 0.1\). The applicability of suppressing the dissipative conductivity using a bias DC, as pioneered by Gurevich~\cite{2014_Gurevich}, is now found to extend to higher frequencies.
The inset of Fig.~\ref{fig5} (d) shows the bias dependence of $\sigma_1$ at $\hbar\omega/\Delta_0 = 0.01$ for different temperatures. The reduction rate is maximized around $T/T_c \sim 0.3$ at this frequency.

As discussed in Section~II, the theory applies unmodified to a semi-infinite material composed of an extreme type II superconductor [Figure~\ref{fig1}(b)]. In particular, the calculation of the surface resistance under a bias DC magnetic field \(B_0\), assuming \(T \ll T_c\) and \(B_0 \ll B_c\), is straightforward because, under these conditions, the dependence of \(\lambda\) on \(B_0\), often referred to as the nonlinear Meissner effect, is relatively weak. See Appendix~\ref{appendix} for more details.

\subsection{Imaginary part of complex conductivity: $\sigma_2$} \label{section_sigma2}

\begin{figure}[tb]
   \begin{center}
   \includegraphics[width=0.49\linewidth]{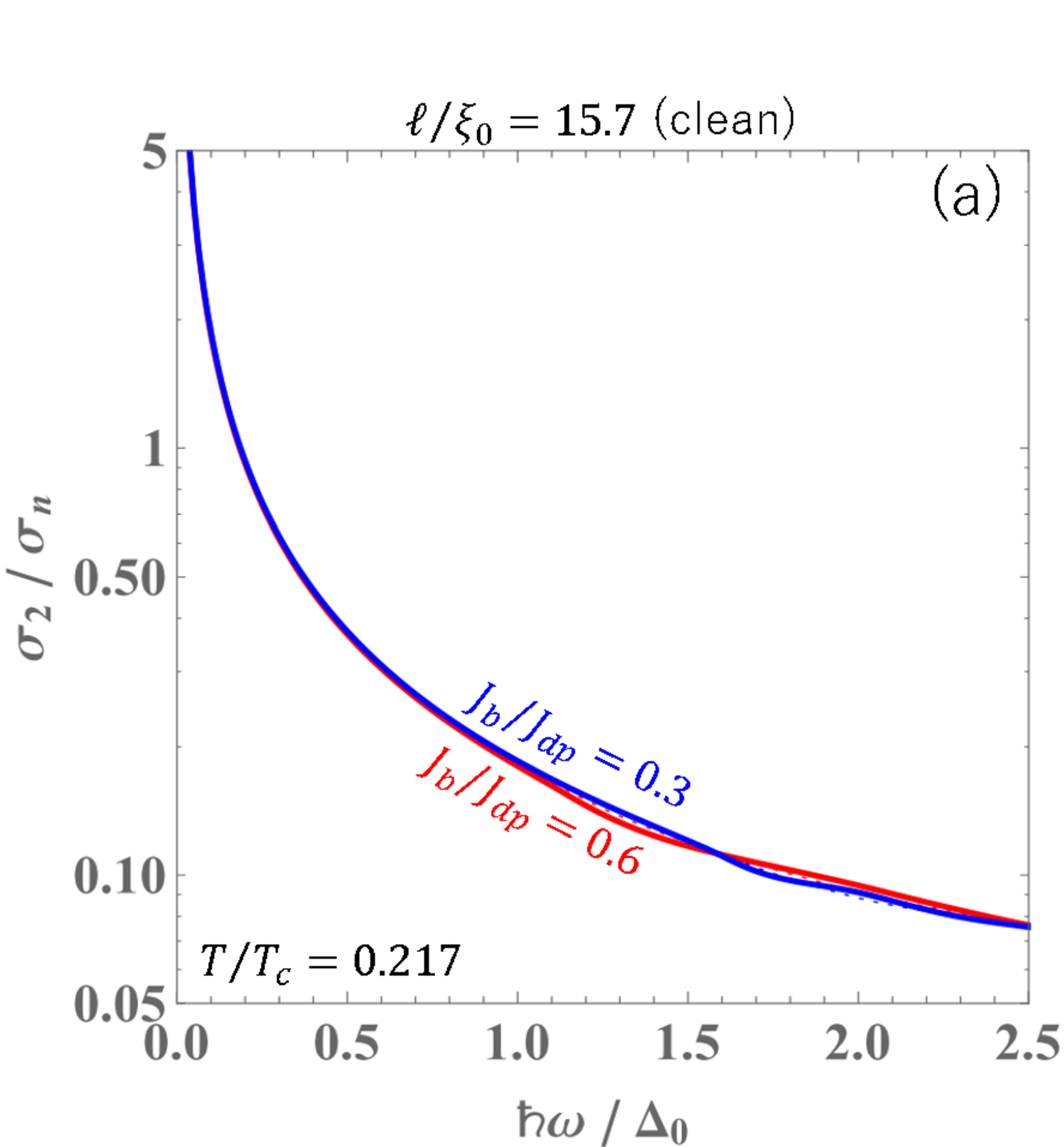}
   \includegraphics[width=0.49\linewidth]{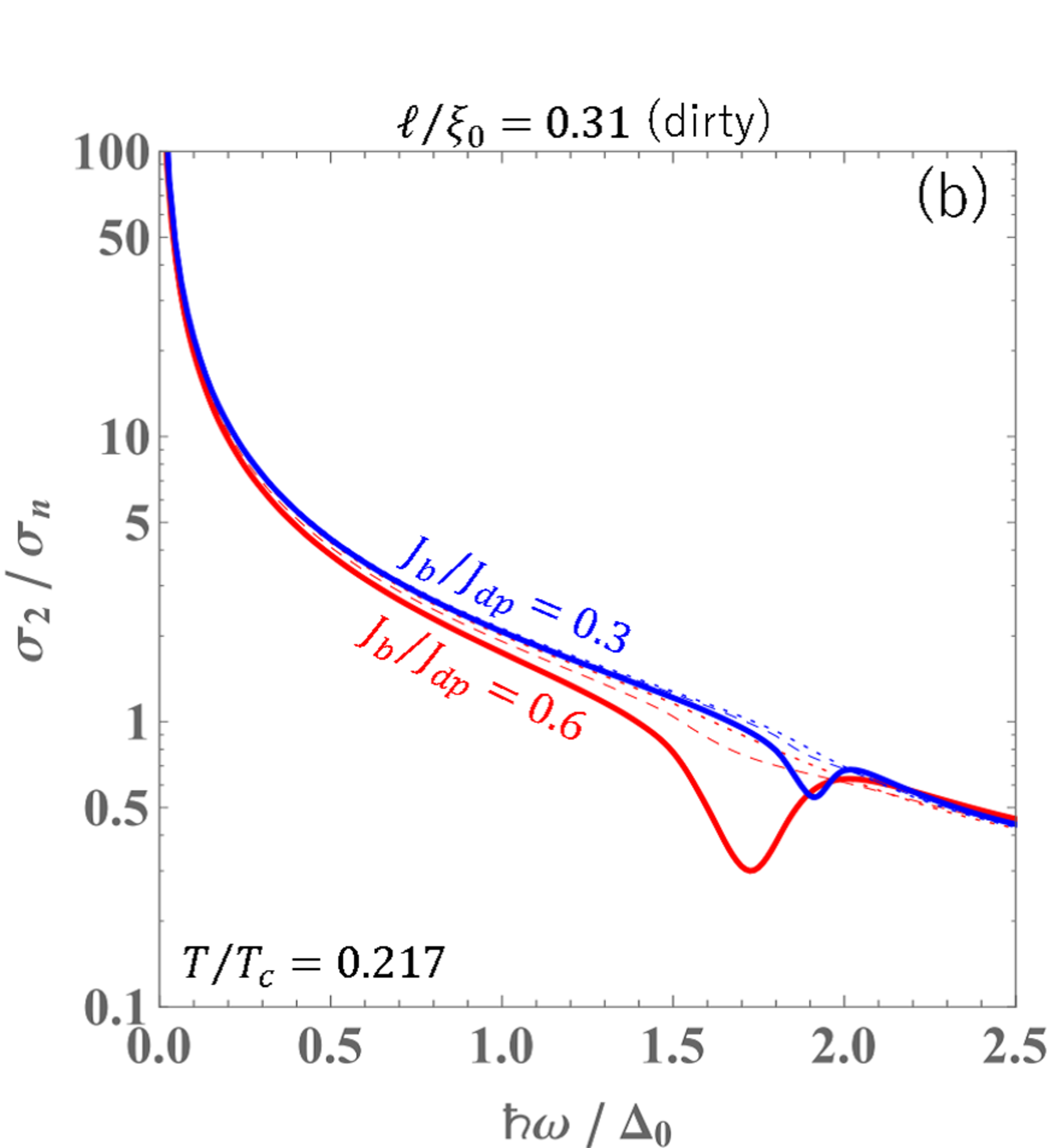}
   \end{center}\vspace{0 cm}
   \caption{
Frequency scan of the imaginary part of the complex conductivity, \(\sigma_2/\sigma_n\), for (a) a clean and (b) a dirty superconductor. 
The dotted and dashed curves represent \(\sigma_2^{(0)}\) and \(\sigma_2^{(0)} + \sigma_2^{\rm (imp)}\), respectively, while the solid curve [\(\sigma_2=\sigma_2^{(0)} + \sigma_2^{\rm (imp)} +\sigma_2^{(\Psi)} \)] includes the contribution from the Higgs mode, \(\sigma_2^{(\Psi)}\). 
The characteristic dip, attributed to the Higgs mode, is evident in the dirty superconductor. 
   }\label{fig6}
\end{figure}

Next, we examine the imaginary part of the complex conductivity, \(\sigma_2\). Figure~\ref{fig6} displays the frequency dependence of \(\sigma_2\) at various strengths of bias DC. In dirty superconductors, as shown in Fig.~\ref{fig6}(b), the characteristic signature of the Higgs mode contribution \(\sigma_2^{(\Psi)}\) is evident as a notable dip at \(\hbar \omega/\Delta_0 \simeq 2\). Ref.~\cite{Nakamura} also theoretically calculated this dip in \(\sigma_2\) using the perturbative formulation for dirty-limit superconductors developed by Moor, which treats both the bias DC and the oscillating current as perturbations, and experimentally observed the dip. Our calculation, which treats the bias DC nonperturbatively, finds a similar dip and shows that the dip grows and shifts to lower frequencies as the bias DC increases. Furthermore, our theoretical framework allows us to consider not only the dirty limit but also any impurity concentration, demonstrating that there is no clear dip in the clean case. This phenomenon is attributed to the Galilean invariance of clean superconductors, as discussed in Section~\ref{section_sigma1}.

\begin{figure}[tb]
   \begin{center}
   \includegraphics[height=0.53\linewidth]{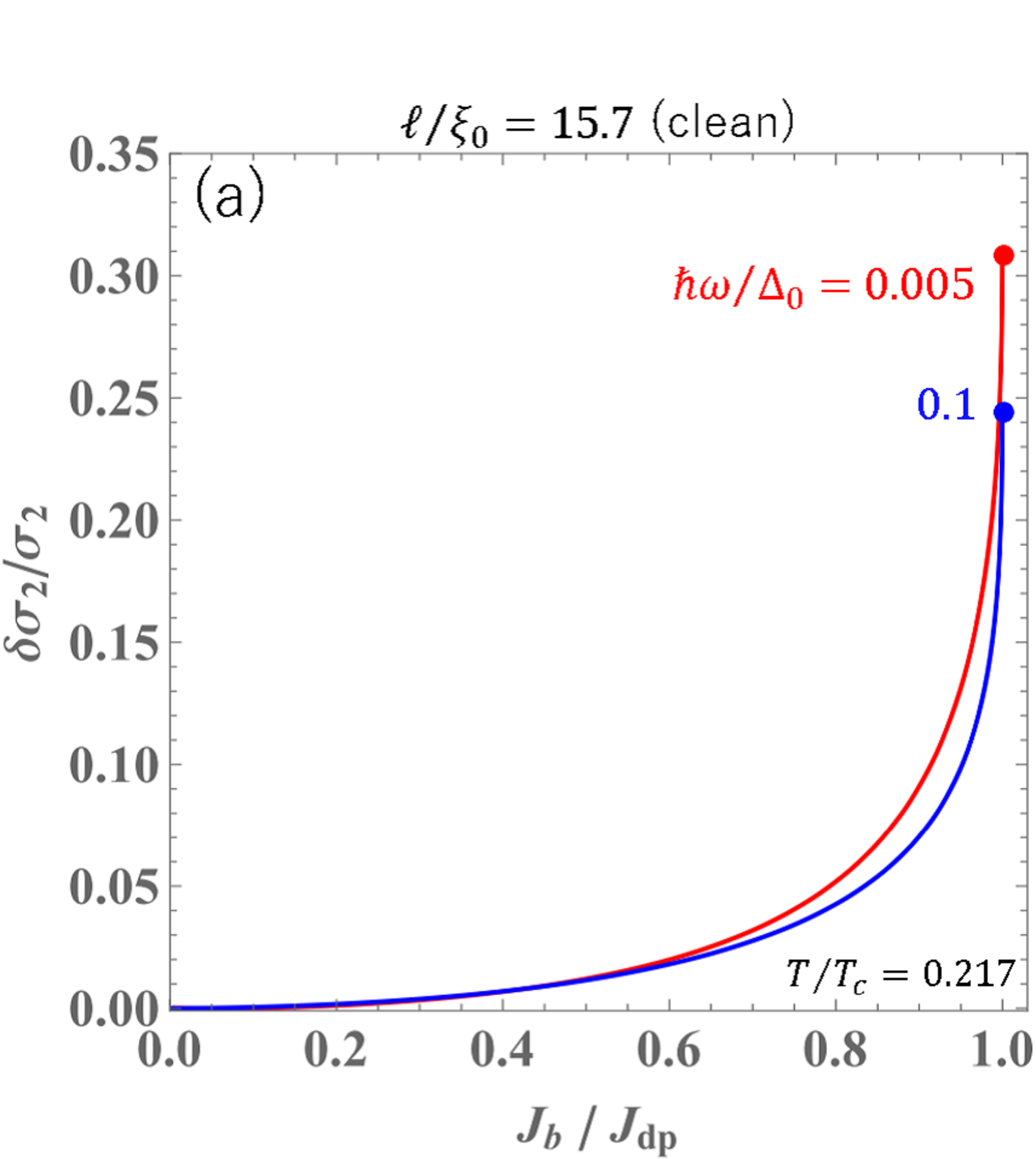}
   \includegraphics[height=0.53\linewidth]{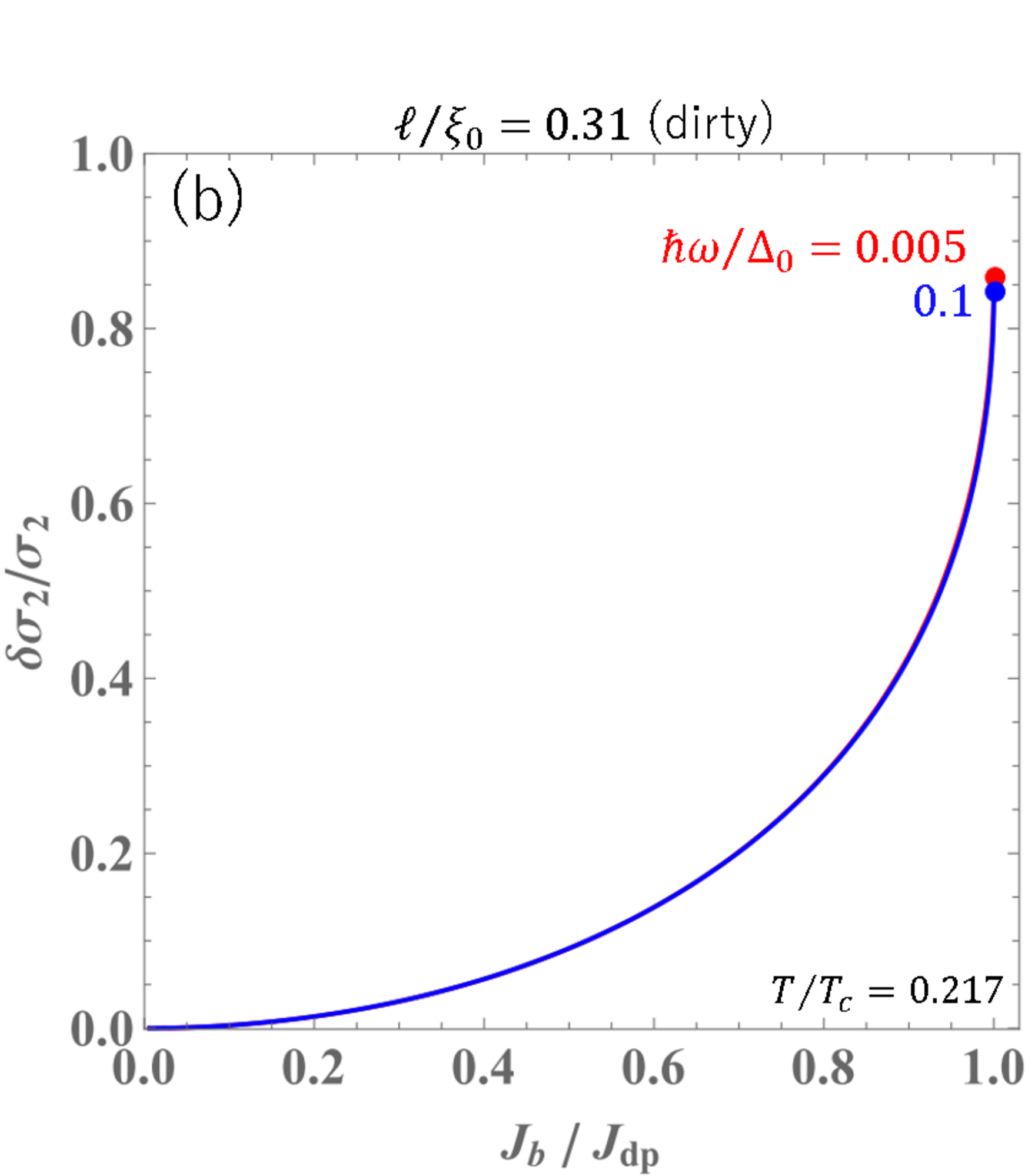}
   \includegraphics[width=0.52\linewidth]{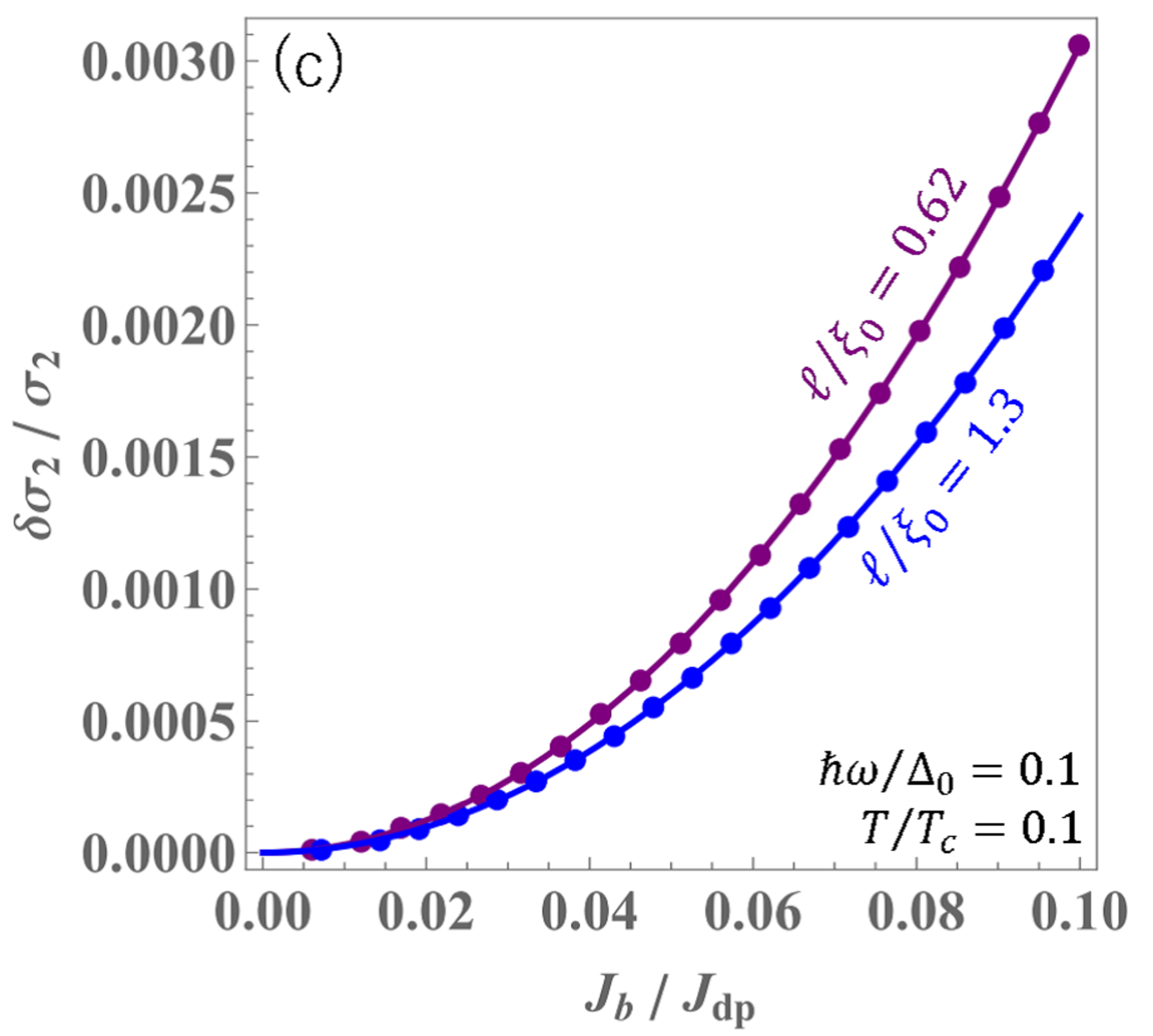}
   \includegraphics[width=0.46\linewidth]{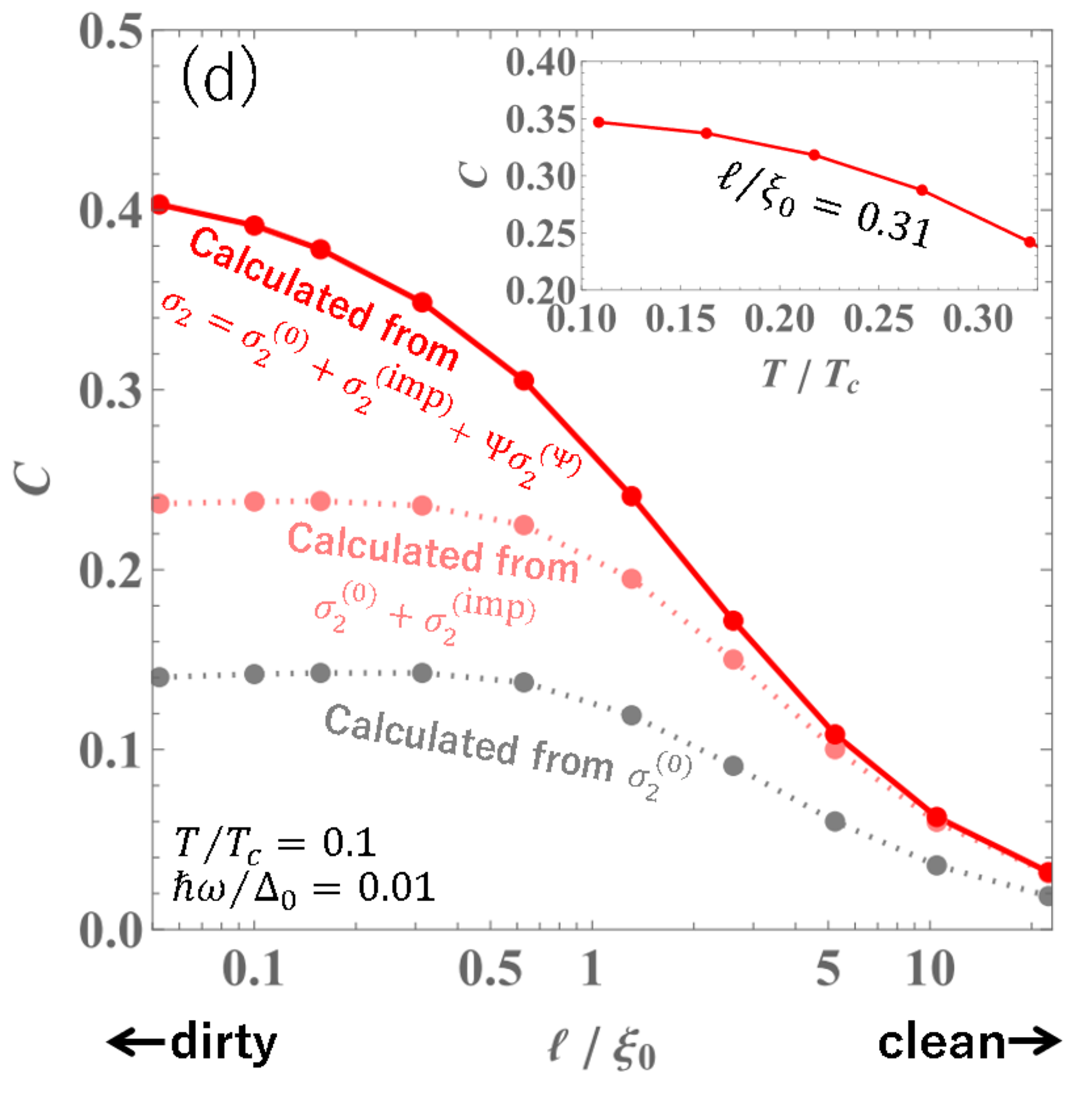}
   \end{center}\vspace{0 cm}
   \caption{
Bias DC dependence of the imaginary part of the complex conductivity \(\delta\sigma_2/\sigma_2 = [\sigma_2(0) - \sigma_2(J_b)]/\sigma_2(0)\), representing the (normalized) bias-dependent kinetic inductance, calculated at \(T/T_c = 0.217\). This is shown for (a) clean and (b) dirty superconductors across various frequencies. The red and blue curves correspond to \(\hbar \omega /\Delta_0 = 0.005\) and \(0.1\), respectively. The blue and red dots indicate the values at the depairing current density.
(c) The normalized bias-dependent kinetic inductance at $J_b/J_{\rm dp} < 0.1$. The purple and blue dots represent the numerical results calculated from the theory. The solid blue and purple curves are the fits to the numerical results using the approximate formula given by Eq.~(\ref{nonlinear_kinetic}).
(d) Bias-dependent kinetic inductance coefficient \(C\) as a function of the mean free path. The red curve includes both contributions from the Higgs mode and impurity scattering self-energy corrections. The pink and gray curves represent incomplete calculations, omitting one or both of these corrections.
The inset shows $C$ as functions of $T$ for a moderately dirty case ($\ell/\xi_0=0.31$). 
   }\label{fig7}
\end{figure}

Figures~\ref{fig7}(a) and (b) illustrate the bias DC dependence of \(\delta \sigma_2/\sigma_2 = [\sigma_2(J_b=0) - \sigma_2(J_b)]/\sigma_2(J_b=0)\), which represents the (normalized) bias-dependent kinetic inductance and is proportional to the bias-dependent frequency shift of resonators. In both the clean and dirty cases, the frequency dependence is small within the range \(\hbar \omega/\Delta_0 = 0.01\) to \(0.1\) and is negligible (less than \(1\%\)) in small bias DC regions.

Let us focus on small bias DC regions. The bias-dependent kinetic inductance for \(J_b/J_{\rm dp} \lesssim 0.1\) can be expressed as  
\begin{eqnarray}
\frac{\delta \sigma_2}{\sigma_2} = C \left( \frac{J_b}{J_{\rm dp}} \right)^2 . \label{nonlinear_kinetic}
\end{eqnarray}
Since our calculation is based on the Keldysh-Eilenberger formalism of nonequilibrium superconductivity, which provides a rigorous theoretical framework to calculate bias-dependent kinetic inductance while accounting for nonequilibrium effects across a range of impurity concentrations from dirty to clean cases, we can determine the bias-dependent kinetic inductance coefficient \(C\) by combining our calculation results shown in Figs.~\ref{fig7}(a) and (b) with Eq.~(\ref{nonlinear_kinetic}) or \(\sigma_2(J_b) = \sigma_2(0) [1 - C (J_b/J_{\rm dp})^2]\).
For example, Figure~\ref{fig7}(c) shows \(\delta \sigma_2/\sigma_2\) for different mean free paths calculated for the small bias region \(J_b/J_{\rm dp} < 0.1\). The two curves correspond to \(C = 0.306\) (purple) and \(0.241\) (blue). Repeating this procedure, we obtain the mean free path dependence and temperature dependence of the coefficient \(C\).

Figure~\ref{fig7}(d) shows the bias-dependent kinetic inductance coefficient \(C\) as a function of the mean free path \(\ell/\xi_0 = (\pi\Delta_0/2\gamma)\). The inset shows the $T$ dependence of $C$ for a  moderately dirty case ($\ell/\xi_0=0.31$).
In the dirty-limit regime (\(\ell/\xi_0 \ll 1\)), our calculation, which includes both the contribution of the impurity scattering self-energy corrections \(\sigma_2^{\rm (imp)}\) and that of the Higgs mode \(\sigma_2^{(\Psi)}\), yields \(C \sim 0.4\). Conversely, if we drop the contributions of \(\sigma_2^{\rm (imp)}\) and \(\sigma_2^{(\Psi)}\), we obtain \(C \simeq 0.14\). In the clean-limit regime (\(\ell/\xi_0 \gg 1\)), the coefficient becomes much smaller than in dirty superconductors and exhibits a tiny bias-dependence for \(J_b/J_{\rm dp} < 0.1\).

In previous studies~\cite{Clem_Kogan, 2020_Kubo_2}, the bias-dependent kinetic inductance for dirty superconductors has been calculated based on the oscillating and frozen superfluid density (\(n_s\)) scenarios. The oscillating \(n_s\) scenario assumes \(n_s\) instantly adjusts to the time variance of the oscillating electromagnetic field, whereas the frozen \(n_s\) scenario assumes \(n_s\) cannot respond to the current's time dependence. Some experimental results seem to support the oscillating \(n_s\) scenario (see e.g., Ref.~\cite{Frasca}). By combining these phenomenological scenarios with the equilibrium Usadel equation, analytical formulas for \(C\) at \(T \to 0\) are obtained~\cite{2020_Kubo_2}. According to these formulas, \(C \simeq 0.41\) for the oscillating \(n_s\) scenario and \(C \simeq 0.14\) for the frozen \(n_s\) scenario. Other authors~\cite{2020_Semenov} also obtained \(C \simeq 0.14\) for the DC biased system without considering the Higgs mode contribution, coinciding with the value from the frozen \(n_s\) scenario in Ref.~\cite{2020_Kubo_2}.

Interestingly, \(C = 0.41\) for the oscillating \(n_s\) scenario coincides with our calculation based on the Keldysh-Eilenberger formalism of nonequilibrium superconductivity [see Fig.~\ref{fig7} (d)], where the Higgs mode necessarily contributes to the kinetic inductance under a bias DC. Considering that the Higgs mode is essentially an oscillation of \(n_s\), the oscillating \(n_s\) scenario can be viewed as a phenomenological implementation of the Higgs mode's effect on the kinetic inductance calculation, now for the first time justified by the theory of nonequilibrium superconductivity. Conversely, \(C = 0.14\) for the frozen \(n_s\) scenario matches the incorrect calculation where the contributions from \(\sigma_2^{\rm (imp)}\) and \(\sigma_2^{(\Psi)}\) are omitted. Therefore, the frozen \(n_s\) scenario cannot be theoretically justified. Moreover, as \(\omega\) increases and approaches the resonance frequency of the Higgs mode (\(\simeq 2\Delta\)), the contribution from the Higgs mode, essentially an oscillation of \(n_s\), grows (see Fig.~\ref{fig6}), and the frozen \(n_s\) scenario is not realized even at higher frequencies.

\section{Discussion} 

Starting from the well-established Keldysh-Eilenberger formalism of nonequilibrium superconductivity, we derived the complex conductivity formula for a superconductor under a bias DC (a similar formulation, applicable not only to local but also to nonlocal electrodynamics, has already been presented in Ref.~\cite{Jujo}). This formula includes contributions from the impurity-scattering self-energy correction \(\sigma_{1,2}^{\rm (imp)}\) and the Higgs mode \(\sigma_{1,2}^{(\Psi)}\), which are now recognized as essential for accurate complex conductivity calculations~\cite{Moor, Jujo, Nakamura, Rainer_Sauls}. 
We then calculated the complex conductivity \(\sigma = \sigma_1 + i\sigma_2\) under a bias DC for various impurity concentrations, photon frequencies, and bias intensities.

\subsection{$\sigma_1$}

As shown in Fig.~\ref{fig4}, including the Higgs mode and impurity scattering self-energy corrections is essential for accurately calculating \(\sigma_1\) under a bias DC. In dirty superconductors, the Higgs mode significantly impacts the results, creating a peak at \(\hbar \omega/\Delta_0 \simeq 2\), as highlighted in previous studies~\cite{Moor, Jujo, Nakamura}.

These corrections notably influence \(\sigma_1\) at lower frequencies (\(\hbar\omega \ll \Delta_0\)) in both clean and dirty superconductors. As shown in Fig.~\ref{fig5}, they significantly alter the bias DC-dependent \(\sigma_1\). Contrary to previous beliefs that \(\sigma_1\) reduction diminishes with increasing \(\omega\), our results show it actually grows with \(\omega\). The low-frequency surface resistance \(R_s\) also shows similar behavior, suggesting that bias DC can tune dissipation in superconducting devices across a broader frequency range than previously expected~\cite{2014_Gurevich, 2020_Kubo_1}.
Future work will involve a detailed exploration of the parameter space $(T, \omega, \ell, J_b)$, tailored to specific device applications.

\subsection{$\sigma_2$}

Accurately calculating the imaginary part of the complex conductivity $\sigma_2$ under a bias DC also requires including the Higgs mode and self-energy corrections, especially in dirty superconductors. As shown in Fig.~\ref{fig6}, the Higgs mode creates a distinct dip at \(\hbar \omega/\Delta_0 \simeq 2\), as highlighted in previous studies~\cite{Moor, Jujo, Nakamura}.

The importance of these corrections is evident in the calculations of the bias-dependent normalized change in the imaginary part of the complex conductivity or the bias-dependent kinetic inductance, \(\delta\sigma_2/\sigma_2\). Fig.~\ref{fig7}(d) focuses on small current regions to determine the coefficient \(C\) for the formula \(\delta\sigma_2/\sigma_2 = C \times (J_b/J_{\rm dp})^2\), applicable when \(J_b/J_{\rm dp} \ll 1\).
Our calculations based on the Keldysh-Eilenberger formalism of nonequilibrium superconductivity yield \(C \simeq 0.41\) in the dirty limit, coinciding with previous studies based on the oscillating \(n_s\) (slow experiment) scenario. Since the Higgs mode is essentially an oscillation of \(n_s\), this agreement is reasonable and justifies the oscillating \(n_s\) scenario through the robust theory of nonequilibrium superconductivity.

Using this theoretical framework, we can calculate \(C\) for any impurity concentration at any temperature, as shown in Fig.~\ref{fig7}(d). Future work will involve comparing the theoretical and experimental results for the temperature and mean free path dependence of \(C\).


\begin{acknowledgments}
I am sincerely grateful to everyone who generously supported my extended paternity leave, which spanned three years. Your support allowed me to prioritize and cherish invaluable time with my family~\cite{ikuji}. This work originated from discussions with Akira Miyazaki in February 2024, right after my leave ended, which reminded me of an email exchange with Anatoly Volkov about the Higgs mode just before my leave. These interactions inspired me to tackle this topic. I also appreciate the insightful discussions with Hiroki Kutsuma regarding the physics of kinetic inductance during his short stay at KEK in March 2024. This work was supported by JSPS KAKENHI Grants No. JP17KK0100 and Toray Science Foundation Grants No. 19-6004.
\end{acknowledgments}

\appendix

\section{Surface resistance in a semi-infinite superconductor under a bias DC magnetic field} \label{appendix}

\begin{figure}[tb]
   \begin{center}
   \includegraphics[height=0.53\linewidth]{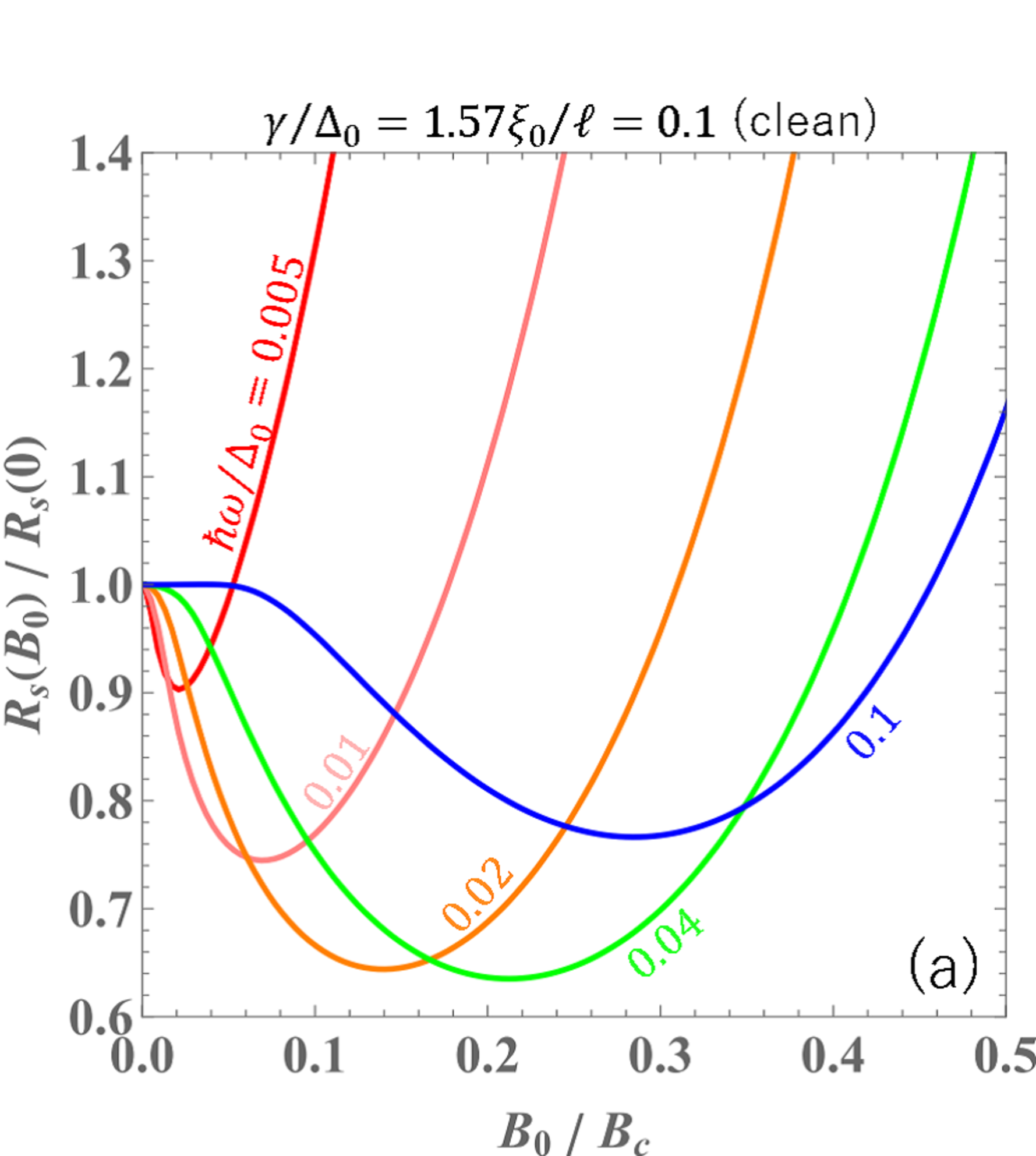}
   \includegraphics[height=0.53\linewidth]{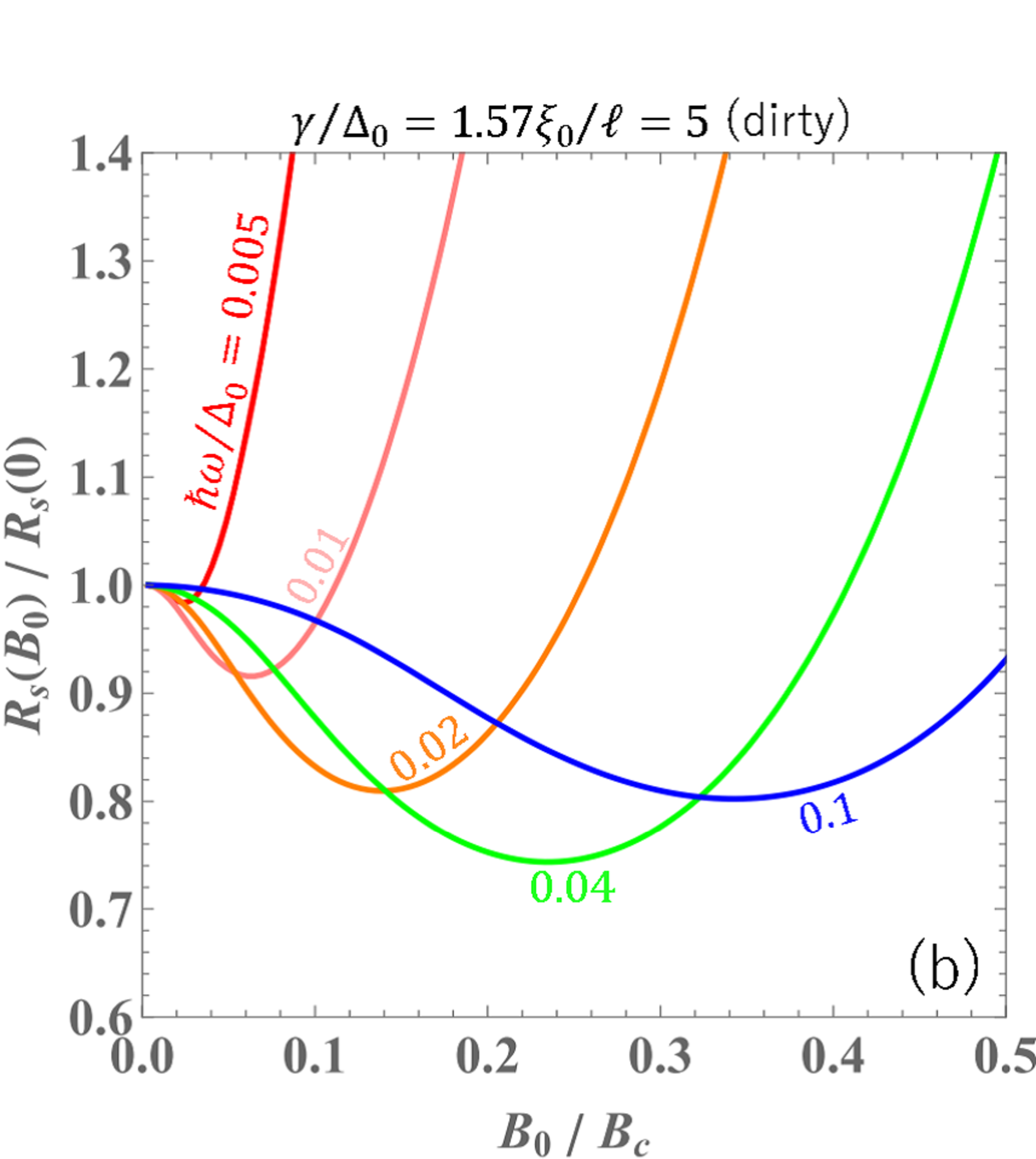}
   \end{center}\vspace{0 cm}
   \caption{
Surface resistances \(R_s\) of semi-infinite superconductors plotted as functions of the bias DC magnetic field \(B_0/B_c\), for a range of frequencies (\(\hbar \omega/\Delta_0 = 0.005\) to \(0.1\)). These calculations were performed at a reduced temperature of \(T/T_c = 0.217\).
   }\label{fig8}
\end{figure}

As discussed in Section~II, the theory applies unmodified to a semi-infinite material composed of an extreme type II superconductor. Consider Figure~\ref{fig1}(b), where the DC magnetic field \(B_0\) is applied parallel to the \(z\)-axis. For simplicity, we assume \(T \ll T_c\) and \(B_0 \ll B_c\), conditions under which the dependence of \(\lambda\) on \(B_0\), known as the nonlinear Meissner effect, is relatively weak. In this scenario, the Meissner effect is represented by \(B(x) = B_0 e^{-x/\lambda}\), and the distribution of the superfluid momentum \(q_b(x)\) is given by \(q_b(x) = q_b(0) e^{-x/\lambda}\). Here, \(q_b(0)\) is related to \(B_0\) through the relationship \(q_b(0)/q_0 = (\sqrt{6}/\pi) (\lambda/\lambda_0) (B_0/B_c)\), where \(B_c = \sqrt{\mu_0 N_0 \Delta_0^2}\) is the zero-temperature thermodynamic critical field, \(\lambda_0^{-2} = (2/3) \mu_0 N_0 e^2 v_f^2\) is the clean-limit zero-temperature London depth, and \(\lambda\) is the London depth for the considered superconductor.
According to BCS theory, at $T\ll T_c$, \(\lambda/\lambda_0\) can be expressed as \(\sqrt{(\gamma/\Delta_0)/(\pi/2 - h(\gamma))}\), where \(h\) is defined as \(\cos^{-1}(\gamma/\Delta_0)/\sqrt{1 - (\gamma/\Delta_0)^2}\) for \(\gamma/\Delta_0 < 1\) and \(\cosh^{-1}(\gamma/\Delta_0)/\sqrt{(\gamma/\Delta_0)^2 - 1}\) for \(\gamma/\Delta_0 > 1\) (see, e.g., Refs.~\cite{2017_Gurevich_review, 2022_Kubo}).

This leads to the following expressions for the surface resistance:
\begin{eqnarray}
R_s = \mu_0^2 \omega^2 \lambda^2 \int_0^{\infty} dx \, \sigma_1(x) e^{-2x/\lambda} , 
\end{eqnarray}
or
\begin{eqnarray}
\frac{R_s(B_0)}{R_s(0)} = \frac{\int_0^{q_b(0)} dq_b \, q_b \sigma_1(q_b)}{\frac{1}{2} q_b(0)^2 \sigma_1(0)} .
\end{eqnarray}
Figure~\ref{fig8} presents the bias-dependent surface resistance for both moderately clean and moderately dirty cases, calculated across various frequencies. The results are similar to those shown in Figs.~\ref{fig5}(c) and (d).
To calculate the cases of higher DC fields, it is necessary to take into account the nonlinear Meissner effect in a bulk superconductor~\cite{2021_Kubo, Wave_Sauls}.


\begin{thebibliography}{99}



\bibitem{Moor}
A. Moor, A. F. Volkov, and K. B. Efetov, 
Amplitude Higgs Mode and Admittance in Superconductors with a Moving Condensate,
Phys. Rev. Lett. {\bf 118}, 047001 (2017).

\bibitem{Jujo}
T. Jujo, 
Surface Resistance and Amplitude Mode under Uniform and Static External Field in Conventional Superconductors, 
J. Phys. Soc. Jpn. {\bf 91}, 074711 (2022). 

\bibitem{Nakamura}
S. Nakamura, Y. Iida, Y. Murotani, R. Matsunaga, H. Terai, and R. Shimano, 
Infrared Activation of the Higgs Mode by Supercurrent Injection in Superconducting NbN,
Phys. Rev. Lett. {\bf 122}, 257001 (2019). 

\bibitem{Shimano_review}
R. Shimano and N. Tsuji, 
Higgs mode in superconductors, 
Annu. Rev. Condens. Matter Phys. {\bf 11}, 103 (2020).


\bibitem{2012_Zmuidzinas}
J. Zmuidzinas, 
Superconducting Microresonators: Physics and Applications, 
Annu. Rev. Condens. Matter Phys. {\bf 3}, 169 (2012).

\bibitem{2017_Gurevich_review}
A. Gurevich, 
Theory of RF superconductivity for resonant cavities, 
Supercond. Sci. Technol. {\bf 30}, 034004 (2017). 


\bibitem{Tsuji_review}
N. Tsuji, I. Danshita, S. Tsuchiya,
Higgs and Nambu-Goldstone modes in condensed matter physics,
in {\it Encyclopedia of Condensed Matter Physics (Second Edition)}, edited by Tapash Chakraborty (Academic Press, Canmbridge, Massachusetts, 2024), p. 174. 

\bibitem{Anderson}
P. W. Anderson, 
Random-Phase Approximation in the Theory of Superconductivity, 
Phys. Rev. {\bf 112}, 1900 (1958). 

\bibitem{Volkov_Kogan}
A. F. Volkov and S. M. Kogan,
Collisionless relaxation of the energy gap in superconductors,
Soviet Journal of Experimental and Theoretical Physics {\bf 38}, 1018 (1974). 

\bibitem{2013_Matsunaga}
R. Matsunaga, Y. I. Hamada, K. Makise, Y. Uzawa, H. Terai, Z. Wang, and R. Shimano, 
Higgs Amplitude Mode in the BCS Superconductors ${\rm Nb}_{1-x} {\rm Ti}_x {\rm N}$ Induced by Terahertz Pulse Excitation, 
Phys. Rev. Lett. {\bf 111}, 057002 (2013). 

\bibitem{2014_Matsunaga}
R. Matsunaga, N. Tsuji, H. Fujita, A. Sugioka, K. Makise, Y. Uzawa, H. Terai, Z. Wang, H. Aoki, and R. Shimano, 
Light-induced collective pseudospin precession resonating with Higgs mode in a superconductor, 
Science {\bf 345}, 1145 (2014). 

\bibitem{Tsuji_Aoki}
N. Tsuji, and H. Aoki, 
Theory of Anderson pseudo spin resonance with Higgs mode in superconductors, 
Phys. Rev. B {\bf 92}, 064508 (2015).

\bibitem{2018_Jujo}
T. Jujo, 
Quasiclassical Theory on Third-Harmonic Generation in Conventional Superconductors with Paramagnetic Impurities, 
J. Phys. Soc. Jpn. {\bf 87}, 024704 (2018).

\bibitem{Silaev}
M. Silaev,
Nonlinear electromagnetic response and Higgs-mode excitation in BCS superconductors with impurities,  
Phys. Rev. B {\bf 99}, 224511 (2019).








\bibitem{2014_Gurevich}
A. Gurevich, 
Reduction of Dissipative Nonlinear Conductivity of Superconductors by Static and Microwave Magnetic Fields, 
Phys. Rev. Lett. {\bf 113}, 087001 (2014). 

\bibitem{2020_Kubo_1}
T. Kubo, 
Weak-field dissipative conductivity of a dirty superconductor with Dynes subgap states under a dc bias current up to the depairing current density, 
Phys. Rev. Res. {\bf 2}, 013302 (2020). 

\bibitem{Clem_Kogan}
J. R. Clem and V. G. Kogan, 
Kinetic impedance and depairing in thin and narrow superconducting films,
Phys. Rev. B {\bf 86}, 174521 (2012). 

\bibitem{2020_Semenov}
A.V. Semenov, I. A. Devyatov, M. P. Westig, and T. M. Klapwijk, 
Effect of Microwaves on Superconductors for Kinetic Inductance Detection and Parametric Amplification, 
Phys. Rev. Applied {\bf 13}, 024079 (2020). 

\bibitem{2020_Kubo_2}
T. Kubo,
Superfluid flow in disordered superconductors with Dynes pair-breaking scattering: Depairing current, kinetic inductance, and superheating field, 
Phys. Rev. Res. {\bf 2}, 033203 (2020). 
(Note: The formula for $C_{sm}$ for $T=0$ is correct, but the numerical value 0.544 is a typo and must be 0.409.)


\bibitem{Rainer_Sauls}
D. Rainer and J. A. Sauls, 
Strong coupling theory of superconductivity, 
in {\it Superconductivity: From Basic Physics to the Latest Developments}, 
edited by P. N. Butcher and Y. Lu (World Scientific, Singapore, 1995), p. 45.

\bibitem{Kopnin}
N. B. Kopnin, {\it Theory of Nonequilibrium Superconductivity} (Oxford University Press, 2001).


\bibitem{2019_Kubo_Gurevich} 
T. Kubo and A. Gurevich, 
Field-dependent nonlinear surface resistance and its optimization by surface nanostructuring in superconductors,
Phys. Rev. B {\bf 100}, 064522 (2019). 

\bibitem{2021_Kubo}
T. Kubo, 
Superheating fields of semi-infinite superconductors and layered superconductors in the diffusive limit: structural optimization based on the microscopic theory, 
Supercond. Sci. Technol. {\bf 34}, 045006 (2021). 

\bibitem{Wave_Sauls}
V. Ngampruetikorn, J. A. Sauls, 
Effect of inhomogeneous surface disorder on the superheating field of superconducting RF cavities
Physical Review Research {\bf 1}, 012015 (2019). 


\bibitem{1963_Maki}
K.  Maki, 
On persistent currents in a superconducting alloy I, 
Prog. Theor. Phys. {\bf 29}, 10 (1963). 


\bibitem{KL}
M. Yu Kupriyanov and V. F. Lukichev, 
Temperature dependence of pair-breaking current in superconductors, 
Fizika Nizkikh Temperatur {\bf 6}, 445 (1980).  

\bibitem{Lin_Gurevich}
F. Pei-Jen and A. Gurevich, 
Effect of impurities on the superheating field of type-II superconductors, 
Phys. Rev. B  {\bf 85}, 054513 (2012). 


\bibitem{2022_Kubo}
T. Kubo, 
Effects of Nonmagnetic Impurities and Subgap States on the Kinetic Inductance, Complex Conductivity, Quality Factor, and Depairing Current Density, 
Phys. Rev. Applied  {\bf 17}, 014018 (2022). 


\bibitem{Zimmermann}
W. Zimmermann, E. Brandt, M. Bauer, E. Seider, and L. Genzel, 
Optical conductivity of BCS superconductors with arbitrary purity,  
Physica C {\bf 183}, 99 (1991).

\bibitem{Herman}
F. Herman and R. Hlubina, 
Microwave response of superconductors that obey local electrodynamics
Phys. Rev. B {\bf 104}, 094519 (2021).

\bibitem{Ueki}
H. Ueki, M. Zarea, J. A. Sauls, 
Electromagnetic Response of Superconducting RF Cavities, 
JPS Conf. Proc. {\bf 38}, 011068 (2023). 


\bibitem{Frasca}
S. Frasca, B. Korzh, M. Colangelo, D. Zhu, A. E. Lita, J. P. Allmaras, E. E. Wollman, V. B. Verma, A. E. Dane, E. Ramirez, A. D. Beyer, S. W. Nam, A. G. Kozorezov, M. D. Shaw, and K. K. Berggren, 
Determining the depairing current in superconducting nanowire single-photon detectors
Phys. Rev. B  {\bf 100}, 054520 (2019). 



\bibitem{ikuji}
T. Kubo, 
An Encouraging of Paternity Leave: A Physicist Who Has Become a Stay-at-Home Dad in New York, 
KASOKUKI, {\bf 20}, 50 (2023). 








\end{thebibliography}
\end{document}